\UseRawInputEncoding
\pdfoutput=1

\documentclass[lettersize,journal]{IEEEtran}
\usepackage{ifpdf}
\usepackage{cite}
\usepackage{algpseudocode}
\usepackage{hyperref}
\usepackage{comment}
\usepackage{amsmath}
\usepackage{amsthm,amssymb,amsfonts}
\usepackage[ruled,linesnumbered]{algorithm2e}

\usepackage{array}
\usepackage{comment}
\usepackage{url}
\usepackage{color,xcolor}
\usepackage{graphicx}
\usepackage{caption}
\usepackage{subfigure}
\usepackage{float} 
\usepackage{multirow}
\usepackage{stfloats}
\usepackage{bm}
\usepackage{subfigure}

\begin{document}

\title{
Reinforcement Learning Based Robust Volt/Var Control in Active Distribution Networks With Imprecisely Known Delay
}

\author{
Hong Cheng,
Huan Luo,
Zhi Liu,~\IEEEmembership{Senior Member,~IEEE},
Wei Sun,~\IEEEmembership{Senior Member,~IEEE},
Weitao Li, {Member,~IEEE},
Qiyue Li,~\IEEEmembership{Senior Member,~IEEE}

\thanks{H. Cheng, H. Luo, W. Sun, W. Li and Q. Li are with Hefei University of Technology, and Engineering Technology Research Center of Industrial Automation Anhui Province, Hefei, Anhui 230009, China. (email: chenghong@mail.hfut.edu.cn, luohuan@mail.hfut.edu.cn, wsun@hfut.edu.cn, wtli@hfut.edu.cn, liqiyue@mail.ustc.edu.cn)}

\thanks{Z. Liu is with The University of Electro-Communications, Tokyo, Japan. (email: liu@ieee.org)}
\thanks{Corresponding author: Qiyue Li.}
}

\maketitle

\begin{abstract}

Active distribution networks (ADNs) incorporating massive photovoltaic (PV) devices encounter challenges of rapid voltage fluctuations and potential violations. 
Due to the fluctuation and intermittency of PV generation, the state gap, arising from time-inconsistent states and exacerbated by imprecisely known system delays, significantly impacts the accuracy of voltage control.
This paper addresses this challenge by introducing a framework for delay adaptive Volt/Var control (VVC) in the presence of imprecisely known system delays to regulate the reactive power of PV inverters. The proposed approach formulates the voltage control, based on predicted system operation states, as a robust VVC problem. 
It employs sample selection from the state prediction interval to promptly identify the worst-performing system operation state. Furthermore, we leverage the decentralized partially observable Markov decision process (Dec-POMDP) to reformulate the robust VVC problem. We design Multiple Policy Networks and employ Multiple Policy Networks and Reward Shaping-based Multi-agent Twin Delayed Deep Deterministic Policy Gradient (MPNRS-MATD3) algorithm to efficiently address and solve the Dec-POMDP model-based problem. Simulation results show the delay adaption characteristic of our proposed framework, and the MPNRS-MATD3 outperforms other multi-agent reinforcement learning algorithms in robust voltage control.



\end{abstract}

\begin{IEEEkeywords}
Active Distribution Networks, Robust Volt/Var Control, Imprecisely Known Delay, Multi-agent Reinforcement Learning
\end{IEEEkeywords}

%
\IEEEpeerreviewmaketitle

\section{Introduction}
%
%
%
%


To achieve carbon peaking and carbon neutrality goals, the construction of new power systems with high penetration of renewable energy is required. Among them, establishing safe and efficient active distribution networks (ADNs) which consist of a large amount of renewable energy is a key \cite{abad2020PV}.
Recently, the photovoltaic (PV) penetration rate in ADNs continues to escalate, and the overvoltage hazards caused by PV power backpropagation are gradually becoming a prominent problem \cite{zhao2017network}. To suppress system overvoltage, the IEEE 1547 standard for the first time allows distributed small-capacity PV inverters to participate in voltage control of the distribution network by outputting reactive power \cite{generation2020ieee}. 

Compared to conventional voltage control equipments, which operate on a slower timescale (mostly minutes level) \cite{Jafari2018}, inverters can quickly respond to voltage fluctuations and reduce network power loss by reactive power compensation \cite{Review2019Sun}. Generally, by considering the system state of ADNs (such as load active/reactive power, PV power generation, etc), the controller can devise an optimal scheme for the reactive power of PV inverters. When using a centralized control framework, the controller needs to collect all data and make decisions for the entire ADN. Consequently, centralized voltage control incurs substantial computational costs and communication burdens, rendering it unsuitable for large-scale ADNs \cite{Richardson2012}.


Decentralized control has the capability to achieve global voltage control within ADNs with minimal information exchange \cite{Anton2017}. It divides the entire ADN into distinct sub-regions and assigns tasks to multiple controllers so that each one can solve the voltage control problem over a sub-region. For example, \cite{Zhang2023}  proposes a sensitivity-based decentralized control algorithm that adjusts power compensation for PV inverters and battery energy storage systems. It effectively solves regional voltage problems without using network-wide controllable resources. However, in the pursuit of an optimal scheduling scheme, decentralized traditional optimization methods necessitate precise system topological structure and all network parameters. These methods are burdened by extensive iterative processing and prove challenging to apply to large-scale ADNs characterized by dynamic distributed energy and load variations \cite{Safe2020}.

With the development of deep learning and artificial intelligence, deep reinforcement learning (DRL) based decentralized voltage control methods have attracted extensive attention \cite{Liu2021Robust}. For example, 
\cite{Two2021Sun, Wang2020Data} utilize a multi-agent deep deterministic policy gradient (MADDPG) algorithm to solve the voltage control problem, and effectively reduce voltage violations. 
\cite{Attention2021Cao} proposes an attention-enabled multi-agent deep reinforcement learning (MADRL) framework for decentralized Volt/Var control (VVC). Additionally,
\cite{Deep2022Cao} develops a multi-agent soft actor-critic (MASAC) algorithm for scheduling PV inverters in the multiple sub-regions of ADNs, and the algorithm can mitigate the fast voltage violations.

The decentralized voltage control process inherently exhibits a system delay, even reaching 30 seconds \cite{Jain2019Quasi}. The system delay encompasses communication delays from system state measurement to reception by controllers, optimization solving time, communication delay of the dispatch command from controllers to inverters, and inverter response time. 
Since the time-varying nature \cite{Gholami2020Robust}, the system delay for each future operation time step is imprecisely known.
Given the state fluctuation of ADNs, the delay will cause the system state gap, representing a time inconsistency of state between the sampling time and the control time. And it seriously affects the precision of voltage control \cite{Gor2023Distr}.

To mitigate the impact of system delay, \cite{Mean2021Wei} characterize random or uncontrollable factors caused by the volatility and intermittency of PV power generation as uncertainties, treating them as worst-case scenarios denoted in a deterministic form within the input to attain robust voltage control. While these methods enhance operating robustness against uncertainties, the reliance on worst-case scenarios may not accurately reflect real-time system operation states, posing a challenge to the precision of voltage control. For obtaining the precise voltage control strategy in the presence of system delay, \cite{Gorbachev2023Distributed} designs a delay-independent coordinated controller accounting for communication transmission delays within the estimated allowable range. Furthermore, \cite{Xing2022} presents a method capable of achieving precise voltage control within a specific delay range. It is noteworthy that once the delay surpasses a defined range, the aforementioned methods for mitigating the impact of delay become inapplicable.

Predicting future system states has emerged as a predominant approach to mitigate the challenges arising from the system delay. For example, \cite{Sang2023Safety} utilizes the neural network to predict the power output of PV/load. Nevertheless, the deterministic prediction falls short of capturing the intermittent and fluctuating nature of power signals. In addressing this limitation, \cite{Sizing2021} predicts probability models for load consumption and PV resources, establishes scenarios by deriving random values from the probability models, and merges similar scenarios into one class.
The intervals that are generated from the probability models can specify a confidence range to eliminate low‐probable cases. The merging of similar scenarios helps to effectively solve the VVC problem.
However, since the system delay is imprecisely known, when using the delay value with a determined form as the prediction time horizon, the predicted results of the future system operation state cannot accurately reflect the real-time state. The voltage control command based on the predicted results is imprecise.



To address the above issues, we propose a delay adaptive VVC framework that integrates the confidence interval of the future system operation state from the predictor, sample selection, and delay adaptive (DA) method into the general inverter-based VVC solution framework.
The three components of this framework address specific challenges: the inadequacy in comprehensively summarizing system volatility, the computation burden arising from inputting the entire predicted interval, and the difficulty in precisely determining the delay in voltage control processing. 
Regarding the robustness of voltage control, we formulate a robust VVC problem, the solution to it minimizes the overall bus voltage deviation and network power loss under the worst-performing system operation state. The worst-performing state denotes the possible state that corresponds to the maximum post-scheduling objective value at future control time. In this way, the voltage control scheme has robustness in the face of difficult-to-regulate situations such as severe voltage fluctuations.

The proposed Robust VVC problem needs multiple controllers to regulate the PV inverters in sub-regions of ADN. With the controllers' measurements, we employ a decentralized partially observable Markov decision process (Dec-POMDP) \cite{2016POMDPs} to derive the optimal policy.
For effectively solving the problem based on Dec-POMDP model, we design the Multiple Policy Networks and reward shaping-based Multi-agent Twin Delayed Deep Deterministic Policy Gradient (MPNRS-MATD3) algorithm. We enhance the policy networks in the algorithm for the input of sample set of system operation state and refine the reward by reward shaping (RS) mechanism to accelerate the convergence speed during the training process of the algorithm.



The main contributions of this paper are four-fold:
\begin{enumerate}
\item  A delay adaptive VVC framework including the system operation state prediction, sample selection, and DA method is proposed to achieve delay adaptive voltage control.

\item A robust VVC problem considering the worst-performing system operation state is formulated to ensure the robustness of voltage control.

\item A MPNRS-MATD3 algorithm is designed for efficiently solving the robust VVC problem by utilizing a Dec-POMDP model and RS mechanism, enhancing the policy networks.

\item A detailed simulation is conducted to prove the delay adaptive characteristic and robustness of VVC schemes and the superiority of the proposed method.

\end{enumerate}

The rest of the article is organized as follows. Section \ref{DAIMR_voltage_control_Framework} introduces the delay adaptive VVC framework and the Robust VVC problem. In section \ref{MARL_Voltage_Control_Strategy}, we formulate the problem as a Dec-POMDP model and solve the optimization problem with the MPNRS-MATD3 algorithm. Section \ref{sec_experiments} lists the results of the simulation. Finally, section \ref{sec_conclusion} summarizes the whole paper.

\section{Delay Adaptive Inverter-based Multi-Region Voltage Control Framework}
\label{DAIMR_voltage_control_Framework}

\begin{figure}[!htb]
    \centering
    \includegraphics[width=3.5in]{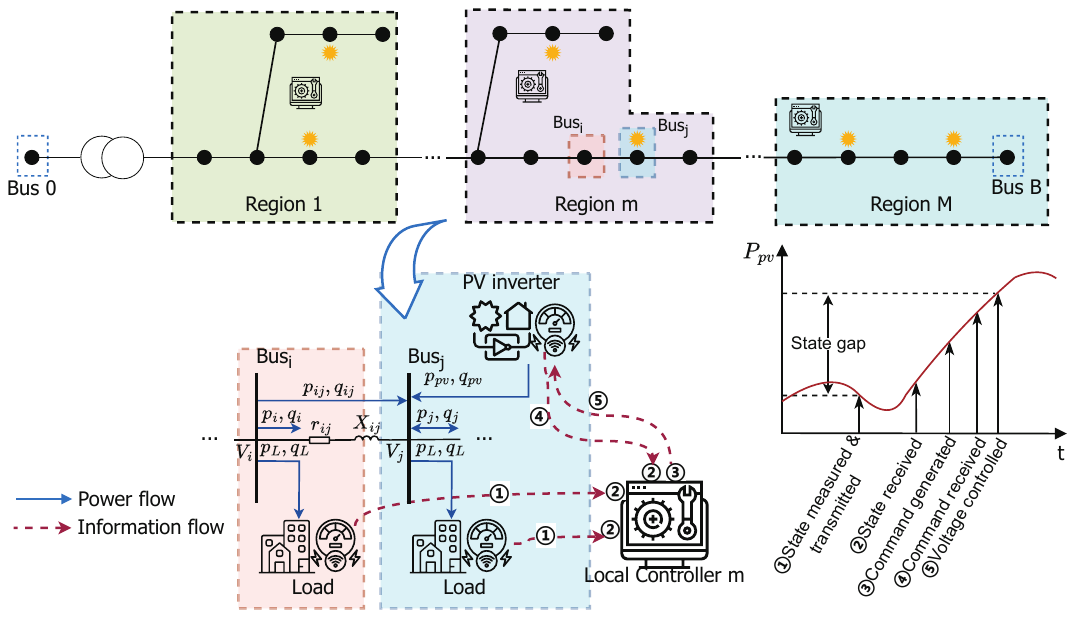}
    \caption{Illustrations of a typical ADN with state gap.}
    \label{fig_ADN_framework}
\end{figure}

As depicted in Fig. \ref{fig_ADN_framework}, we assume a typical ADN with $B$ buses is divided into $M$ regions. Inter-regional information exchange is through the power flow of edge buses. In the region $m$, the set of buses is $B_m$, the set of branches is $E_m$, and the set of PV inverters is $D_m$. The load on bus $i$ absorbs active and reactive power, $p_{i}^{L}$, $q_{i}^{L}$, from the network, and the PV equipment points on bus $j$ inject or absorb active and reactive power, $p_{j}^{PV}$, $q_{j}^{PV}$, into the network via the inverter. 
The branch connecting bus $i$ and $j$ has the resistance, reactance, conductivity, and susceptance, denoted as $r_{ij}$, $x_{ij}$, $g_{ij}$, $b_{ij}$, respectively. According to power flow in ADN, the voltage amplitude and phase angle of the bus $i$, $v_{i}$, $\theta_{i}$, can be obtained. Each region has a local controller, which controls the reactive power of all PV inverters in the region. The system operation state, denoted as $sos_t=\{p_{i,t}^{PV}, p_{i,t}^{L}, q_{i,t}^{L}\mid \forall i\in B\}$, captures the active and reactive power injections and absorptions for all buses at time $t$.

Due to the volatility and intermittency of PV power generation. system delay can introduce a state gap between the time from state sampling to voltage regulation, thereby compromising the precision of voltage control.
This issue can lead to voltage fluctuations and even violations in the ADN. The utilization of probability interval prediction not only effectively addresses the issue of state gaps but also provides a more comprehensive summary of ADN volatility arising from PV generation.

\vspace{-7.5cm}

\begin{figure}[htbp]
    \centering
    \includegraphics[width=3.8in]{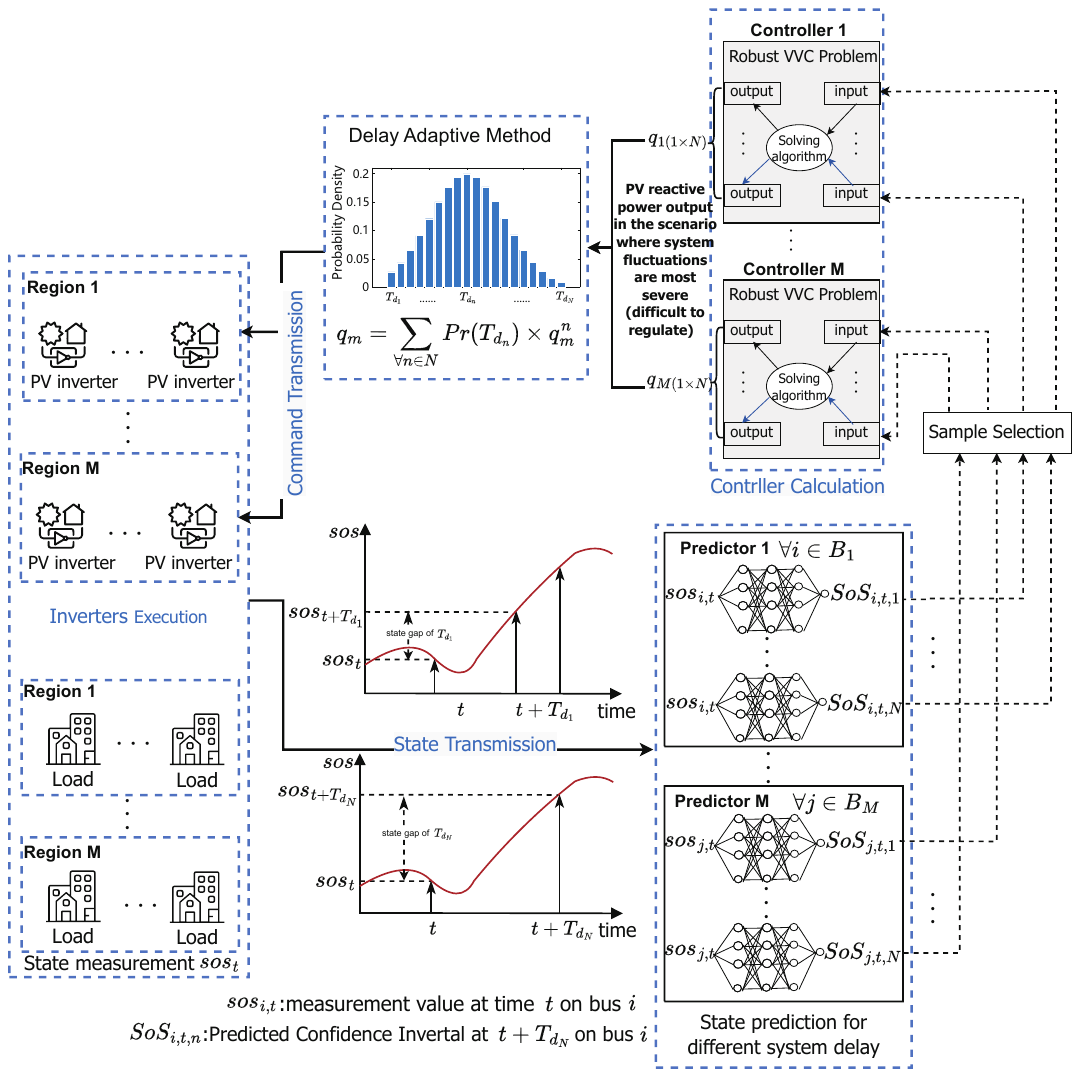}
    \caption{Delay adaptive VVC Framework in ADNs.}
    \label{DAVC_framework}
\end{figure}

To address challenges arising from system delay and state gap issues, we propose a Delay Adaptive VVC framework illustrated in Fig. \ref{DAVC_framework}. This framework encompasses system operation state measurement and prediction, sample selection, controller calculation, and a DA method.
At scheduling time step $t$, we employ the delay value, $T_{d_n}$, as a prediction horizon to estimate the confidence interval of the system operation state, denoted as $SoS_{i,t,n}$, specifically on bus $i$. To account for the imprecisely known delay, we consider multiple possible delay values as prediction horizons, each leading to distinct predictions. 
Subsequently, we use the prediction interval of the system operation state to design the robust VVC problem. By solving this problem, we can determine the optimal solution for PV reactive power, aiming to minimize both system power loss and total voltage deviation in the worst-performing state of system operation.
To efficiently pinpoint the worst-performing system operation state, we classify the case of the estimated state with the same predictive characteristics and apply sample selection to establish a set, $SS_{i,t,n}$, representing the set of real system operation state at the control time.
Finally, due to imprecisely known delay, recognizing that a single delay-corresponding control command may not achieve both delay adaptability and precise voltage control. We utilize the delay probability distribution to design a DA method.

\subsection{System Operation State Prediction }
\label{Delay_Prediction_Model}

According to the historical data of delay, we can determine the system delay range, denoted as $[\underline{T}d, \overline{T}d]$. From this range, we select $N$ possible delay values, forming a set $\{T_{d_1}, ..., T_{d_n}, ..., T_{d_N}\}$. For a specific system delay $T_{d_n}$, we set the prediction horizon as $T_{d_n}$ and assume that the elements of the system operation state follow a Gaussian distribution. 
Employing a confidence level $\delta$, the predictor yields the confidence interval $SoS_{i,t,n}=\{P_{i,t,n}^{PV}, P_{i,t,n}^{L}, Q_{i,t,n}^{L}\}$. This confidence interval encompasses the true value of the system operation state at the control time with a probability of at least $\delta$, a concept illustrated as follows.

\begin{equation} \label{delay_Analysic1}
\setlength\abovedisplayskip{3pt}
\setlength\belowdisplayskip{3pt}
\begin{split}
\rho(p_{i,t,n}^{PV}\in P_{i,t,n}^{PV})\geq\delta,\\
\rho(p_{i,t,n}^{L}\in P_{i,t,n}^{L})\geq\delta,\\
\rho(q_{i,t,n}^{L}\in Q_{i,t,n}^{L})\geq\delta.
\end{split}
\end{equation}

Through the predictor, we can obtain the sets of confidence intervals, ${SoS_{i,t,1},..., SoS_{i,t,n},..., SoS_{i,t,N}}$, corresponding to $N$ delays respectively.

\vspace{-4.5cm}

\subsection{Inverter-based Scheduling Model}
\label{Inverter_based_Scheduling_Model}

We employ the commonly used inverter-based scheduling model \cite{PV2016} to prioritize the provision of reactive power by PV inverters when necessary. Within this model, insufficient reactive power compensation capacity results in a reduction of active power. The inverter's active power is constrained within a specified range. 
Additionally, the reactive power of each inverter is limited to a preset proportion of its apparent power capacity. A non-negative reactive power value signifies injection into the ADN, while conversely, a negative value indicates absorption from the ADN. The constraints of the inverter scheduling model are described as follows:

\begin{equation} \label{power_limits1}
\setlength\abovedisplayskip{3pt}
\setlength\belowdisplayskip{3pt}
\begin{split}
(p_{i,t}^{PV})^{2}+(q_{i,t}^{PV})^{2}\leq(s_{i}^{PV})^{2},
\end{split}
\end{equation}

\begin{equation} \label{power_limits2}
\setlength\abovedisplayskip{3pt}
\setlength\belowdisplayskip{3pt}
\begin{split}
p_{i,min}^{PV}\leq p_{i,t}^{PV}\leq p_{i,max}^{PV},
\end{split}
\end{equation}

\begin{equation} \label{power_limits3}
\setlength\abovedisplayskip{3pt}
\setlength\belowdisplayskip{3pt}
\begin{split}
-\beta s_{i}^{PV}\leq q_{i,t}^{PV}\leq \beta s_{i}^{PV},
\end{split}
\end{equation}
where $p_{i,max}^{PV}$ is maximum active power generation at bus $i$, $\beta$ is inverter reactive power capacity factor.

\subsection{Delay Adaptive Enabled Robust Optimization Problem Formulation}
\label{Problem_Formulation}

To improve the power quality of end users and enable utilities to reduce operational and maintenance costs, the VVC optimization goal for each control region is to minimize overall bus voltage deviation and network power loss. 
Additionally, we define the worst-performing system operation state as the state associated with the maximum post-scheduling objective value, and this state represents the difficult-to-regulate situation in ADN such as severe voltage fluctuations.
To achieve robust voltage control, we solve the VVC problem within this worst-performing system operation state.
Specifically, when the system delay is denoted as $T_{d_n}$ and the operational step is $t$, the formulation of the robust VVC problem is as follows:

\label{system_model}
\begin{equation} \label{Voltage_Control_model1}
\setlength\abovedisplayskip{3pt}
\setlength\belowdisplayskip{3pt}
\begin{split}
\sum_{m\in M}\min_{\substack{q_{i,t,n}^{PV}\\\forall i\in B_m}}\max_{\substack{sos_{i,t,n},\\\forall i\in B_m}}(\lambda_1 \sum_{i\in B_m}f_{i,t,n}^{\triangle v}+\lambda_2\sum_{ij\in E_m}f_{ij,t,n}^{loss}),\\
\forall t\in T, \forall n\in N
\end{split}
\end{equation}

s.t. (\ref{delay_Analysic1}), (\ref{power_limits1}), (\ref{power_limits2}), (\ref{power_limits3})

\begin{equation} \label{Voltage_Control_model2}
\setlength\abovedisplayskip{3pt}
\setlength\belowdisplayskip{3pt}
\begin{split}
sos_{i,t,n}\in SoS_{i,t,n},
\end{split}
\end{equation}

\begin{equation} \label{voltage_deviation}
\setlength\abovedisplayskip{3pt}
\setlength\belowdisplayskip{3pt}
\begin{split}
f_{i,t,n}^{\triangle v}=|v_{i,t,n}-v_{ref}|,
\end{split}
\end{equation}

\begin{equation} \label{Objective_Programming3}
\setlength\abovedisplayskip{3pt}
\setlength\belowdisplayskip{3pt}
\begin{split}
f_{ij,t,n}^{loss}=Re[\frac{(V_{i,t,n}-V_{j,t,n})^2}{r_{ij}-jx_{ij}}],
\end{split}
\end{equation}

\begin{equation} \label{power_flow_equations1}
\setlength\abovedisplayskip{3pt}
\setlength\belowdisplayskip{3pt}
\begin{split}
p_{i,t,n}^{PV}-p_{i,t,n}^{L}=|v_{i,t,n}|\sum_{j\in{B_{m}}}|v_{j,t,n}|[g_{ij}\cos({\theta_{i,t,n}}-{\theta_{j,t,n}})\\+b_{ij}\sin({\theta_{i,t,n}}-{\theta_{j,t,n}})] ,\forall{i}
\end{split}
\end{equation}

\begin{equation} \label{power_flow_equations2}
\setlength\abovedisplayskip{3pt}
\setlength\belowdisplayskip{3pt}
\begin{split}
q_{i,t,n}^{PV}-q_{i,t,n}^{L}=|v_{i,t,n}|\sum_{j\in{B_{m}}}|v_{j,t,n}|[g_{ij}\sin({\theta_{i,t,n}}-{\theta_{j,t,n}})\\-b_{ij}\cos({\theta_{i,t,n}}-{\theta_{j,t,n}})] ,\forall{i}
\end{split}
\end{equation} 
where Equ. (\ref{Voltage_Control_model2}) indicates that the system operation state on the bus $i$ is selected from the confidence interval of the prediction. When the system operation states with the worst performance, the objective function (\ref{Voltage_Control_model1}), connecting Equ. (\ref{voltage_deviation}) and Equ. (\ref{Objective_Programming3}), is to minimize voltage deviation, $f_{t}^{\triangle v}$, and network power loss, $f_{t}^{loss}$. And $\lambda_1$ and $\lambda_2$ are weight coefficients.
Equ. (\ref{voltage_deviation}) is the voltage deviation function. For safety and optimal operation, we need to set reference voltage, $v_{ref}$, and safe voltage range, $\triangle v$. Equ. (\ref{Objective_Programming3}) is utilized to calculate the network power loss, and the power loss of each branch is derived from bus voltages
and branch impedance. Equ. (\ref{power_flow_equations1}) and Equ. (\ref{power_flow_equations2}) are power flow, which can be solved by the Newton-Raphson method.

With the predicted system operation state based on a specified delay $T_{d_n},$ and subsequent solution of the aforementioned robust VVC problem, the command for PV reactive power can be obtained. However, due to the imprecisely known delay, a singular delay-corresponding control command falls short of achieving precise voltage control. Consequently, we propose a DA method. 
Leveraging historical data of system delay, we calculate the mean value, $\mu_{T_d}$, and variance, $\sigma_{T_d}$, assuming a normal distribution for system delay, denoted as $\rho(T_{d_i})=\mathcal{N}(\mu_{T_d},\sigma_{T_d}).$ Employing the system delay probability density function, we assign probabilities to weight the control commands. The delay adaptive control commands for PV inverters can be obtained as follows.

\begin{equation} \label{Adaptive_Control2}
\setlength\abovedisplayskip{3pt}
\setlength\belowdisplayskip{3pt}
\begin{split}
q_{i,t}^{PV}=\sum_{n=1}^N \rho(T_{d_n})\times q_{i,t,n}^{PV},\forall i\in D
\end{split}
\end{equation} 
$q_{i,t,n}^{PV}$ represents the reactive power of the PV inverter at bus $i$ in time step $t$ when the system delay is $T_{d_n}$. Equ. (\ref{Adaptive_Control2}) indicates that the adaptive control command is obtained by weighting $q_{i,t,n}^{PV}$ and the probability values, $\rho(T_{d_n})$, which conforms to statistical laws. 

\subsection{Sample Selection for Robust Voltage Control}
\label{Sample_Selection}

The VVC problem involving the variable PV reactive power is inherently non-convex. The robust optimization problem introduced in Section \ref{Problem_Formulation} is further complicated by the inclusion of variables associated with the system operation state. To streamline the solution of the robust optimization problem, we classify states with the same predictive features and employ sample selection to the confidence interval $SoS$. 
Specifically, we define a set as $SS_{i,t,n}=\{sos^{\mathbb{j}}_{i,t,n}\mid \mathbb{j}=1,2,3\}\in SoS_{i,t,n}$ and use $SS$ to replace the extensive range of system operation state $SoS$ in the Equ. (\ref{Voltage_Control_model2}). Here, $sos^{1}_{i,t,n}$ and $sos^{2}_{i,t,n}$ represent the upper and lower boundaries of $SoS$, and $sos^{3}_{i,t,n}$ denotes the median. This approach can enhance the efficiency of solving the Robust VVC problem.
In pursuit of the solution, we formulate the weighted objective value of region $m$ based on sample selection as:


\begin{equation} \label{Objective_Programming1}
\setlength\abovedisplayskip{3pt}
\setlength\belowdisplayskip{3pt}
\begin{split}
f= (\lambda_1 \sum_{i\in B_m}f_{i,t,n}^{\triangle V}+\lambda_2 \sum_{ij\in E_m}f_{ij,t,n}^{loss}),   
\end{split}
\end{equation}We redefine the objective function (\ref{Voltage_Control_model1}) of robust VVC problem as:

\begin{equation} \label{Objective_Programming2}
\setlength\abovedisplayskip{3pt}
\setlength\belowdisplayskip{3pt}
\begin{split}
f_1= \sum_{m\in M} \min_{\substack{q_{i,t,n}^{PV}\\\forall i\in B_m}}\max_{\substack{\psi_\mathbb{j}\\\forall \mathbb{j}\in \{1,2,3\}}}[\psi_\mathbb{j}\cdot f_\mathbb{j}],
\end{split}
\end{equation}
where $\psi_\mathbb{j}\in\{0,1\}$, $\sum_{\mathbb{j}}\psi_\mathbb{j}=1$ and $f_\mathbb{j}=f(sos^{\mathbb{j}}_{i,t,n})$. By solving the VVC problem with Equ. (\ref{Objective_Programming2}), the robustness of the control is achieved.


\section{MPNRS-MATD3 for Voltage Control}
\label{MARL_Voltage_Control_Strategy}

\begin{figure}[H]
    \centering
    \includegraphics[width=3.8in]{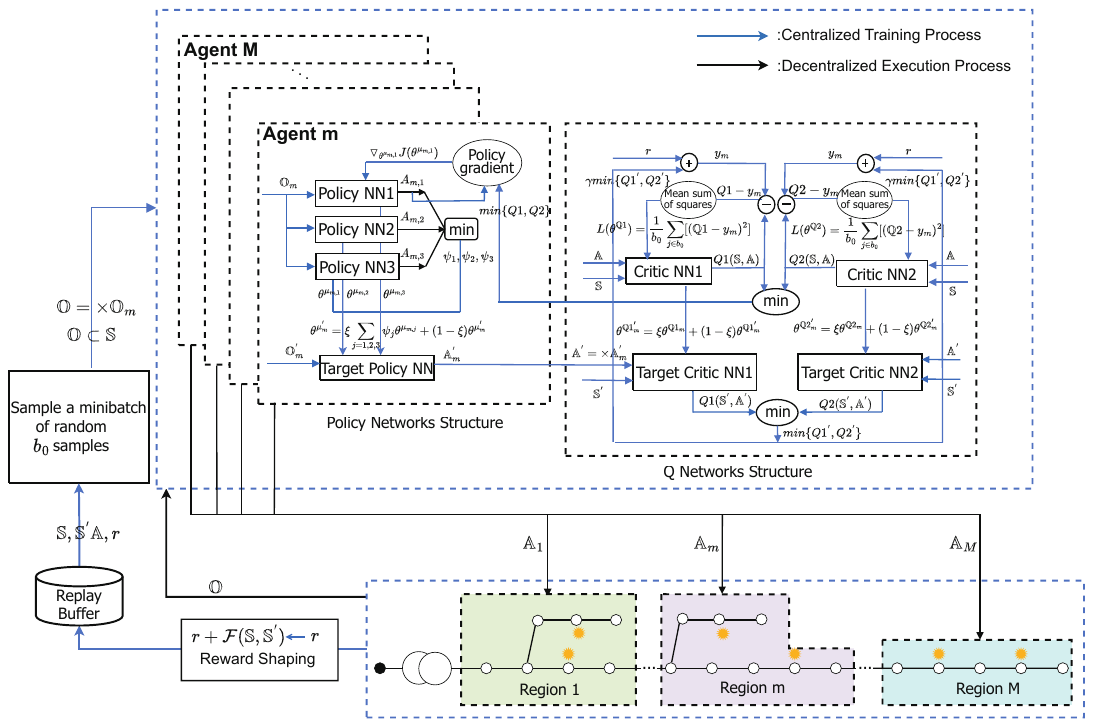}
    \caption{MPNRS-MATD3 Algorithm.}
    \label{MPNRS_MATD3_algorithm}
\end{figure}

When implementing decentralized voltage control in ADN, each region is viewed as an agent, which can conduct regional management of PV inverters. 
Given the inherent volatility and intermittency of PV power generation, coupled with the stochastic nature of user behavior, the system state is significantly time-varying.
With the fact that the measurements of agents are local observations rather than the precise system state, we use Dec-POMDP to reformulate the robust VVC problem.
In the Dec-POMDP model, we can determine the PV reactive power $\{q_{i}^{PV}\mid i\in B\}$ by selecting in action space and obtain the value of $\psi_\mathbb{j}$ through comparison in the reward function.
To solve the robust VVC problem based on the Dec-POMDP model, we propose the MPNRS-MATD3 algorithm as illustrated in Fig. \ref{MPNRS_MATD3_algorithm}.
In this algorithm, to accelerate the convergence speed of the training process, we utilize a potential function in the RS mechanism to establish a new reward function.

\subsection{Problem Reformulation into Dec-POMDP}
\label{Problem_reformulation_Dec-POMDPs}
Dec-POMDP is defined as a tuple $\langle\mathbb{I},\mathbb{S},\mathbb{O},\mathbb{A},\mathbb{P},O,r,\gamma,s_0\rangle$, where $\mathbb{I}=\{1,...,M\}$ is the agents set, $\mathbb{S}$ is the states set, $\mathbb{O}$ is the joint observations set, $\mathbb{A}$ is the joint actions set, $\mathbb{P}:\mathbb{S}\times\mathbb{A}\times\mathbb{S}\rightarrow[0,1]$ is the state probability function, $O:\mathbb{S}\times\mathbb{A}\times\mathbb{O}\rightarrow[0,1]$ is the observation probability function, $r$ is the immediate reward function,  $\gamma\in(0,1)$ is a discount factor, $s_0:\mathbb{S}_0\rightarrow[0,1]$ is the initial state distribution. With the delay value $T_{d_n}$, the robust VVC optimization problem proposed in Section \ref{Problem_Formulation} can be reformulated as a Dec-POMDP model. 
\begin{itemize}
	\item $\mathbb{I}$ is the set of $M$ agents, and located at $M$ regions of ADN, each agent controls a set of PV inverters in one region.
	\item $\mathbb{S}=\sum_{\mathbb{j}}(\psi_\mathbb{j}\times sos^\mathbb{j})\times \mathcal{Q}\times V$, and $sos^\mathbb{j}=\{sos^{\mathbb{j}}_{i,t,n}\mid\forall i\in B \}$ . $\mathcal{Q}=\{q_{i}^{PV}\mid\forall i\in B\}$ is a set of the reactive power of PV inverters from the previous time step. $V=\{(v_{i}, \theta_{i})\mid\forall i\in B\}$ is a set of voltage magnitudes and voltage phases. 
    \item $\mathbb{O}=\{\mathbb{O}_m\mid m\in \mathbb{I}\}$, where $\mathbb{O}_m=\{\mathbb{O}^\mathbb{j}_m\mid  \mathbb{j}=1,2,3\}$, and $\mathbb{O}^\mathbb{j}_m=\{sos^\mathbb{j}_{j,t,n}\mid \forall j \in B_m \}$. We define $\widetilde{\mathbb{O}}_m$ is measured system operation state whitin the region $m$, and considering the effect of system delay, $\widetilde{\mathbb{O}}_m\rightarrow\mathbb{O}_m$ is through the prediction and sample selection illustrated in Section.\ref{DAIMR_voltage_control_Framework}.  
	\item $\mathbb{A}=\{\mathbb{A}_m\mid\forall m\in\mathbb{I}\}$, where $\mathbb{A}_{m}=\sum(\psi_\mathbb{j}\cdot\times \mathbb{A}^\mathbb{j}_{m})$, and $\mathbb{A}^\mathbb{j}_{m}=\{a^{\mathbb{j}}_{i}:-\eta\leq a^{\mathbb{j}}_{i}\leq \eta, \eta>0\mid\forall i\in D_m\}$. With the $\mathbb{O}^\mathbb{j}_m$ as input, the element $a^{\mathbb{j}}_{i}$ in the output $\mathbb{A}^\mathbb{j}_{m}$ represents the ratio of maximum PV reactive power. $\eta$ is the upper boundary of the ratio. In addition, PV reactive power $q_{i}^{PV}=a_{i} [(s_{i}^{PV})^2-(p_{i}^{PV})^2]^{\frac{1}{2}}$.

	\item $\mathbb{P}=Pr(\mathbb{S}_{t+1}\mid\mathbb{S}_{t},\mathbb{A}_{t})$, where $\mathbb{S}_{t+1}\in\chi (\mathbb{S}_t, \mathbb{A}_t)$, $\chi(\bullet)$ is the solution of power flow. If the power flow solution converges, set convergence flag $done=1$; otherwise, $done=0$. 
        \item $O=Pr(\mathbb{O}_{t+1}\mid \mathbb{F}(\mathbb{S}_{t+1},\mathbb{A}_{t}))$, where $\widetilde{\mathbb{O}}_{t+1}=\mathbb{F}(\mathbb{S}_{t+1},\mathbb{A}_{t})$,
        the probability from $\mathbb{S}_{t+1}\rightarrow\widetilde{\mathbb{O}}_{t+1}$ is due to the system change in ADN. That is $Pr(\widetilde{\mathbb{O}}_{t+1}\mid\mathbb{S}_{t+1})=Pr(\mathbb{S}_{t+1})+\mathcal{N}(0,\sum)$, and $\mathcal{N}(0,\sum)$ is the isotropic multivariate Gaussian distribution which is related to the physical properties of sensors. $O=Pr(\mathbb{O}_{t+1}\mid\widetilde{\mathbb{O}}_{t+1})$ is due to the prediction and sample selection.
	
        \item $r$ is established based on the system objective function. Since the objective function is to minimize voltage deviation and network power loss in the worst-performance system operation state, the reward function is modeled as $r_t=\min_{\psi_\mathbb{j}}[-\psi_\mathbb{j}\cdot f_\mathbb{j}]$. 

    \end{itemize} 
    
The value of the objective $f_\mathbb{j}$ is calculated under the set $\mathbb{A}^\mathbb{j}_{m}$. 
By comparing the value of $f_\mathbb{j}$, we can determine the binary variables $\psi_\mathbb{j}$, the global state $\mathbb{S}$, action set $\mathbb{A}$ and reward $r$ for calculating the state value function $V^\pi(\mathbb{S})$ and the state action value function $Q^\pi(\mathbb{S}, \mathbb{A})$. 
In the Dec-POMDP, policy $\pi$, which is measured by $V^\pi(\mathbb{S})$ and $Q^\pi(\mathbb{S}, \mathbb{A})$, represents the probability of the agent taking an action in a specific state. 
The aim of each agent is to identify the optimal policy $\pi^{\ast}$ that maximizes the expected return $R=\sum_{\forall t\in T}\gamma^t r_t$ over the time horizon of an episode $T$, where $\gamma\in[0,1]$ is a discount factor balancing the influence of current reward and future return. A value of $\gamma=0$ emphasizes short-term rewards, while a value of 1 prioritizes long-term returns.

\subsection{Reward Shaping Mechanism}
\label{Reward_Shaping_mechanism}

The principle of RS is to improve reward feedback in the environment by adding additional rewards, thereby making progress in discovering high-reward actions. This helps the algorithm reduce the number of training transitions required, and obtain the optimal policy faster. 
\cite{RS1999policy} proves that by using a new reward function $r+\mathbb{\lambda}\mathcal F$ formed by a potential function $\mathcal F$ with a real-value function $\phi:\mathbb{S}\rightarrow R$, and $\mathbb{\lambda}$ is a weighted parameter for adjusting this shaped item,
the same optimal policy as the original reward function $r$ can be generated. \cite{lu2022reward} proposes that equations with the following form can be used as potential functions, 

\begin{equation} \label{reset_states2}
\setlength\abovedisplayskip{3pt}
\setlength\belowdisplayskip{3pt}
\begin{split}
\mathcal F(\mathbb S_t,\mathbb A_t,t,\mathbb S_{t+1}, \mathbb A_{t+1},t+1)=\gamma\phi(\mathbb S_{t+1}, \mathbb A_{t+1},t+1)\\- \phi(\mathbb S,\mathbb A_t,t),
\forall \mathbb S_t\neq \mathbb{S}_0
\end{split}
\end{equation}
For our multi-agent VVC problem with two objectives, the $\phi(s)$ for each agent is designed as follows,

\begin{equation} \label{reset_states2}
\setlength\abovedisplayskip{3pt}
\setlength\belowdisplayskip{3pt}
\begin{split}
\phi_{m}(s)=[1+\frac{{R_{ll,sum}^{ep}}-{R_{ll,max}^{ep}}(t)}{{R_{ll,max}^{ep}(t)}-{R_{ll,min}^{ep}}(t)}]\cdot0.5\\+[1+\frac{{R_{vd,sum}^{ep}}-{R_{vd,max}^{ep}}(t)}{{R_{vd,max}^{ep}(t)}-{R_{vd,min}^{ep}}(t)}]\cdot0.5, t\neq 0
\end{split}
\end{equation}
where $R_{ll}$ and $R_{vd}$ are the rewards about power loss and voltage deviation. $R_{\cdot, sum}^{ep}$ is the sum of the reward in the current episode, $R_{\cdot, max}^{ep}(t)$ and $R_{\cdot, min}^{ep}(t)$ are the maximum/minimum values of episode reward until now.

\subsection{Dec-POMDP Model-based Robust VVC via MPNRS-MATD3}
\label{voltage_control_MARL}

The fluctuation in ADNs may give rise to rapid voltage violations, posing a challenge to the fast-solving capability of the algorithm. 
Moreover, the wide distribution range of buses in ADNs imposes considerable communication costs between buses.
Hence, for the Dec-POMDP model-based robust VVC problem, agents with cooperative relationships necessitate centralized training and decentralized execution. 
The MATD3 algorithm can be deployed for agents, which can learn the optimal strategy under centralized training and execute optimal actions using only local observations. 
Nevertheless, the solution of our proposed Robust VVC problem requires not only determining the PV reactive power scheme but also identifying the worst-performing system operation state. 
The MATD3 algorithm proves insufficient for solving this problem.
To overcome this limitation, we undertake a redesign of the policy networks, incorporate a reward shaping mechanism into the MATD3 algorithm, and propose the MPNRS-MATD3 algorithm.

\vspace{-7.5cm}

In the MPNRS-MATD3 algorithm, we apply experience replay, target networks, and noise exploration, effectively mitigating the impact of sample correlation on neural network training and enhancing overall optimization performance. 
Each agent possesses a set of neural networks, including an ensemble of policy networks $\{\mu^{\mathbb{j}}_{m}\mid\mathbb{j}=1,2,3\}$, Q Networks $\mathbb{Q}1_m, \mathbb{Q}2_m$, target policy network $\mu_{m}^{'}$, and target Q network $\mathbb{Q}1_{m}^{'}, \mathbb{Q}2_{m}^{'}$, with the parameters of them denoted as $\theta^{\bullet}$. For agent $m$,  its centralized state action functions are $\mathbb{Q}1_{m}^{\pi}(\mathbb{S},\mathbb{A}|\theta^{\mathbb{Q}1_m})$, $\mathbb{Q}2_{m}^{\pi}(\mathbb{S},\mathbb{A}|\theta^{\mathbb{Q}2_m})$ and action exploration is $\mathbb{A}^{\mathbb{j}}_m=\mu(\mathbb{O}^{\mathbb{j}}_m|\theta^{\mu^{\mathbb{j}}_m})+\tau$. $\tau$ is noise to enable the agent to explore the environment. 
Each agent learns the optimal behavior by adjusting its policy parameters $\theta^{\mu^{\mathbb{j}}_m}$ towards maximizing the performance objective $J(\theta^{\mu^{\mathbb{j}}_m})$ that is based on the Q value and illustrated as:

\begin{equation} \label{performance_objective}
\setlength\abovedisplayskip{3pt}
\setlength\belowdisplayskip{3pt}
\begin{split}
J(\theta^{\mu^{\mathbb{j}}_m})=\mathbb{E}_{\mathbb{S}\sim\rho^{\pi},\mathbb{A}\sim\mu}[\sum_{t=0}^{\infty}\gamma^{t}r_t].
\end{split}
\end{equation} The proposed MPNRS-MATD3 algorithm for the Robust VVC problem is interpreted as Algo.\ref{MATD3_Algorithm}. When training MPNRS-MATD3, we define $\mathbb{S}^{'}$ and $\mathbb{A}^{'}$ as the state and action spaces for the next operational time step, and the experience replay buffer is represented as $\mathbb{D}=[\mathbb{S}, \mathbb{S}^{'}, \mathbb{A}, r]$. The algorithm uses two Q estimators with the same structure to avoid the issue of overestimation. The update of the centralized Q network is through the loss function $L(\theta^{\mathbb{Q}1_m})$ and $L(\theta^{\mathbb{Q}2_m})$, which is established by utilizing the time difference error method, the target Q network, and target value $y_m$. By employing the inverse transfer of the loss function value, the parameters of the Q network $\theta^{\mathbb{Q}_m}$ gradually converge toward the target Q network. The parameters of the policy networks $\mu^{\mathbb{j}}_m$ are updated in the direction of the policy gradient $\triangledown_{\theta^{\mu^{\mathbb{j}}_m}}J(\theta^{\mu^{\mathbb{j}}_m})$. 
In addition, with the soft update velocity factor $\xi\in(0,1)$, the parameters of the target policy network and target Q network undergo soft updates by the policy network and Q network. The training procedure of MPNRS-MATD3 is detailed in Algo. \ref{Improved_Training_procedure}.

\begin{algorithm}[t]
\caption{MPNRS-MATD3 Algorithm for Voltage Control}
\label{MATD3_Algorithm}
\LinesNumbered
Initialize experience replays $\mathbb{D}$, system delay $T_{d_n}$, initial state $\mathbb{S}_0$, action exploration process with Gaussian noise $\tau$;\\
	\For{each episode}{	

            Predict the confidence regions of system operation state $SoS_{i,t,n}$ under prediction horizon $T_{d_n}$;\\
            Establish a set by sample selection $SS_{i,t,n}$;\\ 
            Solve power flow and update observation as $\mathbb{O}$;\\
            \For{decision timestep $t\in T$}{
                \For{agent $m\in M$}{
                With the input of observation $\mathbb{O}^{\mathbb{j}}_{m}$, select action $\mathbb{A}^{\mathbb{j}}_{m}$, $\forall \mathbb{j}$; 
                }

                \For{$\mathbb{j}\in \{1,2,3\}$}{
                Execute actions $\{\mathbb{A}^{\mathbb{j}}_1,\mathbb{A}^{\mathbb{j}}_2,...,\mathbb{A}^{\mathbb{j}}_M\}$ in power flow, obtain objective value $f_\mathbb{j}$;
                }

                Compare the size of  $f_1,f_2,f_3$,  determine $\psi_1,\psi_2,\psi_3$,  observe reward $r$, global state $\mathbb{S}$, action $\mathbb{A}$, convergence flag $done$;\\ 
                Utilize the power flow result to obtain new state $\mathbb{S}^{'}$;\\
                \If{$done=1$}{
                \For{replay update frequency $i\in \vartheta$}{
                Stack $[\mathbb{S},\mathbb{S}^{'},\mathbb{A},r]$ in replay buffer $\mathbb{D}$;\\
                }
                Update $\mathbb{S}\leftarrow \mathbb{S}^{'}$;\\
                \For{agent $m\in M$}{
                    Execute MPNRS-MATD3 Training procedure and decay noise $\tau$;  
                }
                }
            }
        }
\end{algorithm}

\begin{algorithm}[t]
	\caption{MPNRS-MATD3 Training Procedure}
	\label{Improved_Training_procedure}
	\LinesNumbered
        Sample a random minibatch of $b_0$ transitions $\{[\mathbb{S},\mathbb{S}^{'},\mathbb{A},r]_j\mid\forall j\in b_0\}$ from $\mathbb{D}$;
        \quad \\
        \For{agent $m=1$ to $M$}{
            Initial Q-networks $\{\mathbb{Q}1_{\theta_m},\mathbb{Q}2_{\theta_m}\}$, a policy network ${\mu_{\theta_m}}$, target networks $\{\mathbb{Q}1^{'}_{\theta_m},\mathbb{Q}2^{'}_{\theta_m}\}$, a target policy network ${\mu^{'}_{\theta_m}}$;
            \quad \\
            $\{\mathbb{Q}1^{'}_{\theta_m}\}\leftarrow\{\mathbb{Q}1_{\theta_m}\}$, 
            $\{\mathbb{Q}2^{'}_{\theta_m}\}\leftarrow \{\mathbb{Q}2_{\theta_m}\}$,
            $\{\mu^{'}_{\theta_m}\leftarrow{\mu_{\theta_m}}\}$,
            \quad \\
            \For{iteration step $1$ to $\mathcal T$}{ 
                    Set $y_m=r+\mathbb{\lambda}\mathcal F(\mathbb{S},\mathbb{S}^{'})+\gamma min\{\mathbb{Q}1_{m}^{\mu^{'}}(\mathbb{S}^{'},\mathbb{A}^{'}),\mathbb{Q}2_{m}^{\mu^{'}}(\mathbb{S}^{'},\mathbb{A}^{'})\}|_{\mathbb{A}_{m}^{'}=\mu_{m}^{'}(\mathbb{O}_m)}$;
                    \quad \\ 
                    Update critic by minimizing the loss:
                    \quad \\ $L(\theta^{\mathbb{Q}1_m})=\frac{1}{b_0}\sum_j[(\mathbb{Q}1_{m}^{\mu}(\mathbb{S},\mathbb{A})-y_m)^2]$,
                    \quad \\ $L(\theta^{\mathbb{Q}2_m})=\frac{1}{b_0}\sum_j[(\mathbb{Q}2_{m}^{\mu}(\mathbb{S},\mathbb{A})-y_m)^2]$;
                    \quad \\
                    Update actor using the sampled policy gradient: 
                    \quad \\ $\triangledown_{\theta^{\mu^{\mathbb{j}}_{m}}}J(\theta^{\mu^{\mathbb{j}}_{m}})=\frac{1}{b_0}\sum_j[\triangledown_{\theta^{\mu^{\mathbb{j}}_{m}}}\mu^{\mathbb{j}}_{m}(\mathbb{A}^{\mathbb{j}}_{m,j}|\mathbb{O}^{\mathbb{j}}_{m,j})\quad\cdot\triangledown_{\mathbb{A}^{\mathbb{j}}_{m,j}}Q1^{\pi}_{m,j}(\mathbb{S}_{j},\mathbb{A}_{j})|{\mathbb{A}^{\mathbb{j}}_{m,j}=\mu^{\mathbb{j}}_{m}(\mathbb{O}^{\mathbb{j}}_m)}]$;\quad$\mathbb{j}\in\{1,2,3\}$
            }
            Update target network parameters of each agent $m$: $\theta^{\mu^{'}_{m}}=\xi\theta^{\mu_{m}}+(1-\xi)\theta^{\mu^{'}_{m}}$,
            $\theta^{\mathbb{Q}1^{'}_{m}}=\xi\theta^{\mathbb{Q}1_m}+(1-\xi)\theta^{\mathbb{Q}1^{'}_{m}}$,
            $\theta^{\mathbb{Q}2^{'}_{m}}=\xi\theta^{\mathbb{Q}2_m}+(1-\xi)\theta^{\mathbb{Q}2^{'}_{m}}$
}
\end{algorithm}

\section{Performance Evaluation} 
\label{sec_experiments}


\subsection{Simulation setup}

\begin{figure*}
	\begin{minipage}[H]{0.55\linewidth}
        \vspace{-11cm}
		\centering
		\includegraphics[width=3.98in]{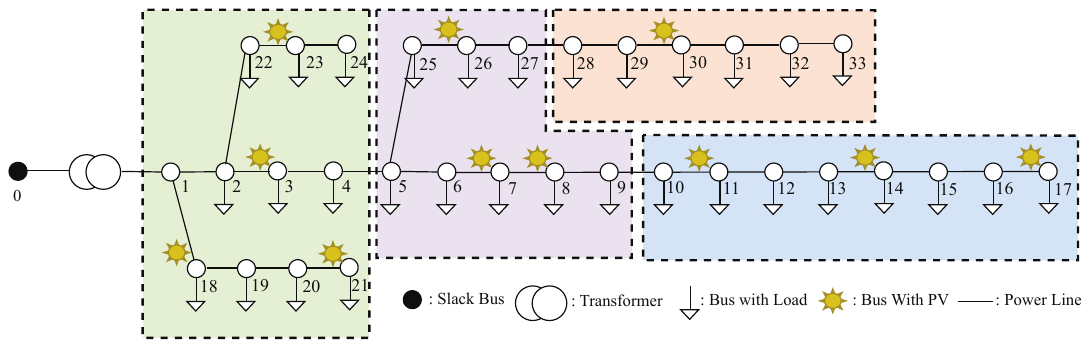}
		\caption{IEEE 33-bus distribution network.}
		\label{fig_IEEE33_Network_Topolopy}
	\end{minipage}
	\begin{minipage}[t]{0.35\linewidth}
        \vspace{-9cm}
		\centering
		\includegraphics[width=2.8in]{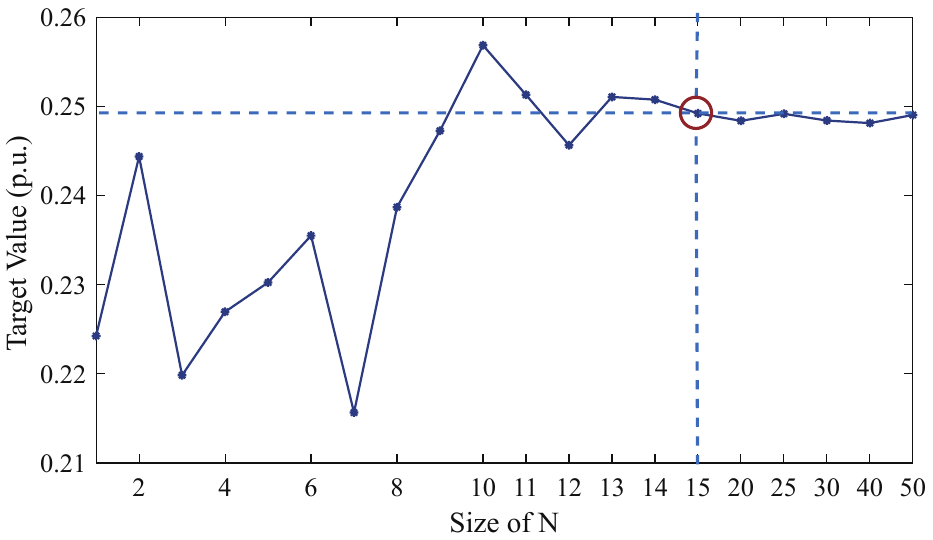}
		\caption{The system performance over different $N$.}
		\label{fig_Determination_N}
	\end{minipage}
\end{figure*}

To evaluate the performance of our proposed scheme, we perform extensive simulations using IEEE 33 distribution test systems. As displayed in Fig. \ref{fig_IEEE33_Network_Topolopy}, the 33-bus network is partitioned into 4 regions, each comprising 1-4 PVs depending on varying regional sizes. All buses except Bus 1 have loads. For the bus equipped with PVs, the inverter apparent power capacity  $s_{i}^{PV}$ is oversized to 120$\%$ of the PV active power capacity $p_{i, max}^{PV}$ to satisfy sufficient reactive power compensation \cite{turitsyn2011options}. All other var resources are assumed to be fixed settings and are not accounted for in this VVC optimization model.
We use the dataset consisting of load and PV generation from \cite{wang2021multi}, which are collected from Elia group\footnote{ \url{https://www.elia.be/en/grid-data/power-generation/solar-pv-power-generation-data.}} and Portuguese electricity consumption\footnote{ \url{https://archive.ics.uci.edu/ml/datasets/ElectricityLoad Diagrams20112014.}}. The test dataset is randomly selected from a 30-minute segment of data in the load and PV generation dataset. The other system parameters are presented in Tab. \ref{table_system_parameter}. 

\begin{table}[!htb]  \centering
\caption{System Parameters Settings}
\label{table_system_parameter}
\small
\begin{tabular}{c|c|c} \hline 
System & \multicolumn{2}{c}{value}\\
\cline{2-3} 
Parameters &33-bus& 141-bus \\ \hline
Number of Buses $B$ & 33&141 \\ \hline
Number of Regions $M$ & 4&9 \\ \hline
Number of PV Inverters $D$ & 11&22 \\ \hline
System Delay $T_d$(s) & \multicolumn{2}{c}{1$\sim$10}  \\ \hline
Prediction Confidence Level $\delta$(\%)  &	\multicolumn{2}{c}{95} \\ \hline
Length of the History of Time Series $t_0$(min) & \multicolumn{2}{c}{5} \\ \hline
Bus Voltage Range(p.u.)  & \multicolumn{2}{c}{[0.95, 1.05]} \\ \hline
Reference Voltage $v_{ref}$(p.u.) & \multicolumn{2}{c}{1} \\ \hline
Inverter Reactive Power Capacity Factor $\beta$ & \multicolumn{2}{c}{0.6} \\ \hline
Weighting Factors $\lambda_1$, $\lambda_2$  & \multicolumn{2}{c}{0.5} \\ \hline
\end{tabular}
\end{table}

\textbf{Implementation details:} The simulations are conducted on a 64-bit PC with Intel Core 6-core 3.7GHz AMD Ryzen 55600X CPU and one NVIDIA GeForce RTX 3060 Ti GPU using Python and Matlab platforms, with the AC Power Flow solved by the PYPOWER \cite{brown2017pypsa} and MATPOWER \cite{zimmerman1997matpower} solvers. The hyperparameters of the MPNRS-MATD3 algorithm are finalized in Tab. \ref{RRS-MATD3_Algorithm_Hyperparameters_Settings}.

\begin{table}[!htb]  \centering
\caption{Hyperparameters Settings of MPNRS-MATD3}
\label{RRS-MATD3_Algorithm_Hyperparameters_Settings}
\small
\begin{tabular}{c|c} \hline 
Hyperparameters & Values   \\ \hline
Discount Factor $\gamma$ & 0.9   \\ \hline
Upper Boundary of & 0.8 \\
 the Ratio $\eta$ &    \\ \hline
Type of Policy Networks & Fully Connected  \\
and Q Networks   &	Neural Networks \\ \hline
Weights Initialization Method & Orthogonal Initialization \\ \hline
Layers of Policy Networks & \multirow{2}{*}{3}\\
$\{\mu_{m,i}|i=1,2,3\}$   & \\ \hline
Layers of Q Networks & \multirow{2}{*}{3}   \\
 $\{\mathbb{Q}i_m|i=1,2\}$ &  \\ \hline
Dimensions of the Hidden & $2d_{\mathbb{O}_m}$, \\
 layers in the Q Networks & $d_{\mathbb{O}_m}$  \\ \hline
Dimensions of the Hidden  & 256,\\ 
layers in the Policy Networks  &  256 \\ \hline
Optimizer & Adam \\ \hline
Learning Rates for the   & \multirow{2}{*}{$5\times10^{-4}$} \\
Policy and Q Networks &  \\ \hline
Reward Discount Factor  $\gamma$ &  0.9 \\ \hline
Soft Update Velocity  & \multirow{2}{*}{0.01} \\ 
Factor $\xi$ & \\ \hline
Capacity of Replay Replay $\mathbb{D}$  & $10^{4}$ \\ \hline
Minibatch Size $b_0$  & 32 \\ \hline
Initial and Minimum Value of  & 0.1, \\
Gaussian Noise $\epsilon$ & 0.02 \\ \hline
Gaussian Noise Decay Steps & 200 \\ \hline
Activation Function & ReLU \\ \hline
Total Decision Timesteps $\mathbb{T}$ & 1000 \\ \hline
Total Iteration Steps $\mathcal T$ & 4000 \\ \hline
\end{tabular}
\end{table}

\begin{figure*}
	\centering
        \subfigure[Load Active Prediction]{
		\begin{minipage}[H]{0.3\linewidth}
			\centering
                \vspace{-4.5cm}
			\includegraphics[width=2in]{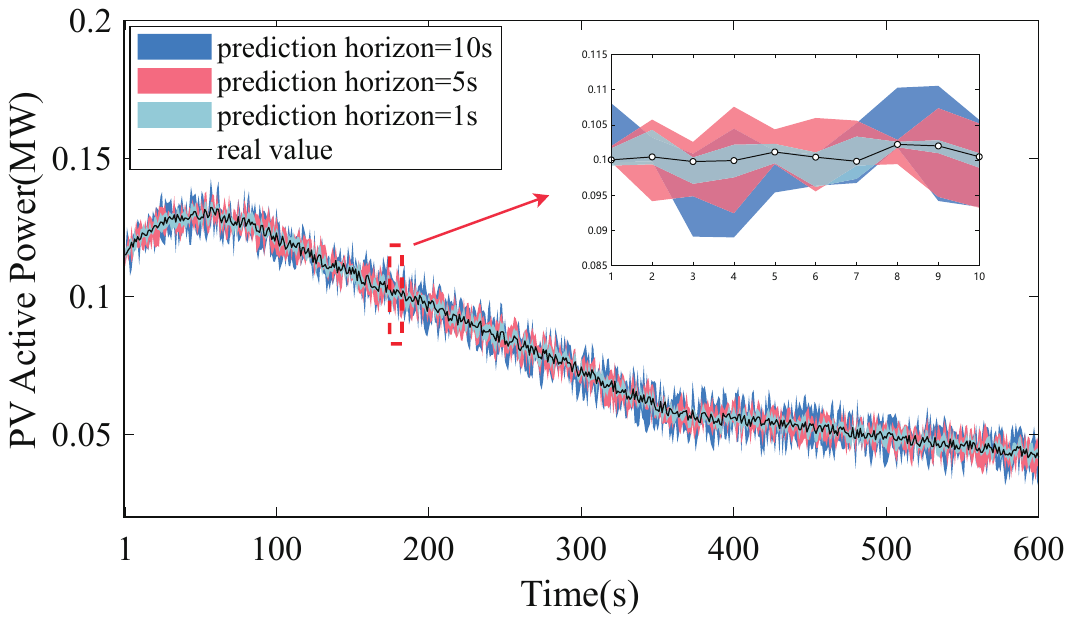}
		\end{minipage}
	}%
        \subfigure[Load Reactive Prediction]{
		\begin{minipage}[H]{0.3\linewidth}
			\centering
                \vspace{-4.5cm}
			\includegraphics[width=2in]{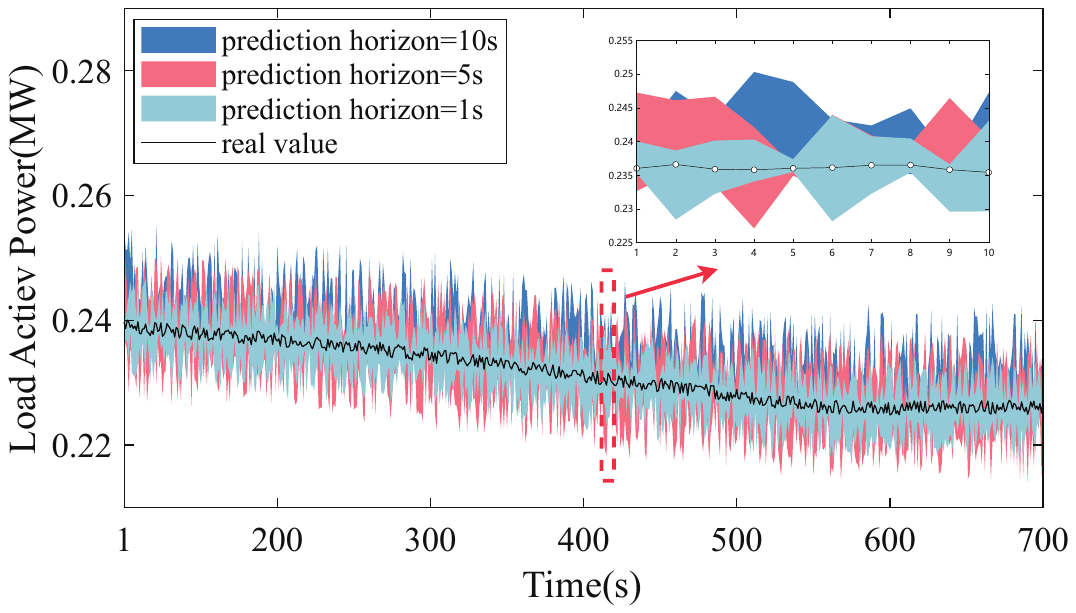}
		\end{minipage}
	}%
	\subfigure[PV Active Prediction]{
		\begin{minipage}[H]{0.3\linewidth}
			\centering
                \vspace{-4.5cm}
			\includegraphics[width=2in]{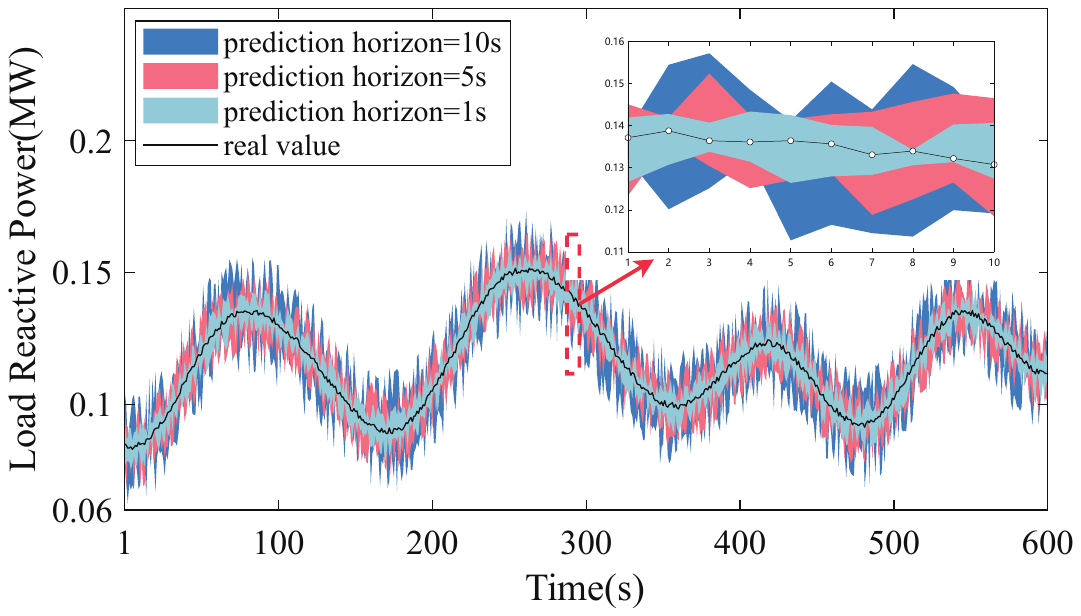}
		\end{minipage}
	}%
	\centering
	\caption{95\% Confidence Region of System Operation State under Different Prediction Horizon}
	\label{fig_prediction}
\end{figure*}

\textbf{System Operation State Prediction:} DeepAR, an encoder-decoder structure-based neural network, is selected as our system operation state predictor\cite{David2020DeepAR}. The learning rate of the DeepAR is set to $10^{-3}$. As indicated in Tab. \ref{table_system_parameter}, given the input of the historical data, the predictor ultimately outputs the 95\% confidence interval of the variable at each prediction time. 
For each bus, the predictor predicts the three elements of $sos$ separately. Fig. \ref{fig_prediction} shows the predicted results of the various prediction horizons (1s, 5s, 10s) on a bus equipped with PV inverter.

\textbf{Comparison Methods:} We select optimal, MADDPG \cite{Liu2021Robust} and MATD3 \cite{Zhang2022MATD3} algorithms as other competitors to evaluate our proposed MPNRS-MATD3. 
To attain the optimal solution, we convert the original non-convex optimization problem into a Second-Order Cone Program (SOCP) through branch flow model phase angle relaxation and solve it with a CPLEX solver.
In recent years, MATD3 and MADDPG have emerged as popular MARL algorithms for voltage control.
Regarding the DA method, we evaluate the following combinations.

\textbf{1) OPT (C1)}: The command of PV reactive power is the optimal solution under the input of real-time data without any system delay.

\textbf{2) OPT+DA (C2)}: The optimal algorithm and DA method are combined to obtain the delay adaptive command of reactive power of all PV inverters.

\textbf{3) OPT+NDA (C3)}: Instead of $N$ predictions in the DA method, we just opt for a singular system delay as the prediction horizon and perform one single $SoS$ prediction, and the optimal algorithm is used to solve the robust VVC problem. Subsequent simulations evaluate the C3, with delays of both 1s and 10s.

\textbf{4) MATD3+DA (C4)}: We use MATD3 to solve the robust VVC problem and the DA method to achieve delay adaptive characteristics of PV reactive power compensation.  

\textbf{5) MADDPG+DA (C5)}: The only difference between \textbf{C4} and \textbf{C5} is that the solving algorithm is MADDPG.

\textbf{$N$ Prediction Horizons}: 
To determine the value of $N$ in the DA method, we use the optimal method to solve the robust VVC problem with test dataset. We calculate the average objective value of the DA method for various $N$ values. 
As shown in Fig. \ref{fig_Determination_N}, when $N>15$, the objective value hardly changes, indicating that 15 is the threshold for the number of possible delay values. And the following test results are consistently obtained with $N=15$.

\subsection{Delay Adaptive Performance}
\label{DAC}
\begin{figure}
	\centering
        \subfigure[Without control]{
        \label{busvoltagerand}
		\begin{minipage}[t]{0.46\linewidth}
			\centering
                 \vspace{-4.5cm}
			\includegraphics[width=1.8in]{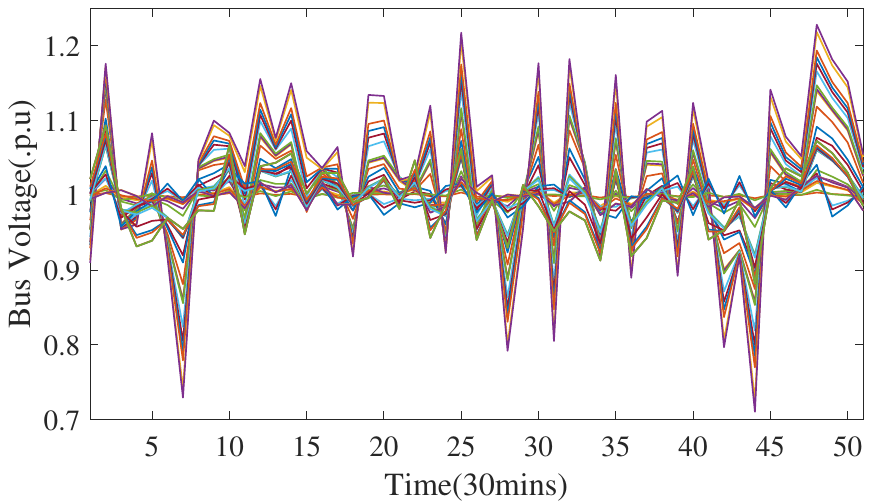}
		\end{minipage}
	}
        \subfigure[C3 (System Delay=10s)]{
        \label{busvoltageNDAC10}
		\begin{minipage}[t]{0.46\linewidth}
			\centering
                 \vspace{-4.5cm}
			\includegraphics[width=1.8in]{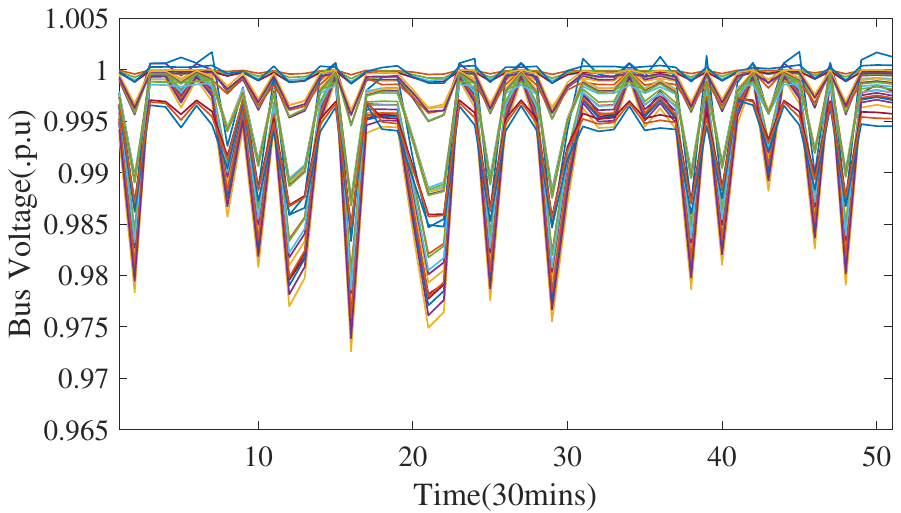}
		\end{minipage}
	}\\%
        \subfigure[C3 (System Delay=1s)]{
        \label{busvoltageNDAC1}
		\begin{minipage}[t]{0.46\linewidth}
			\centering
                 \vspace{-4.5cm}
			\includegraphics[width=1.8in]{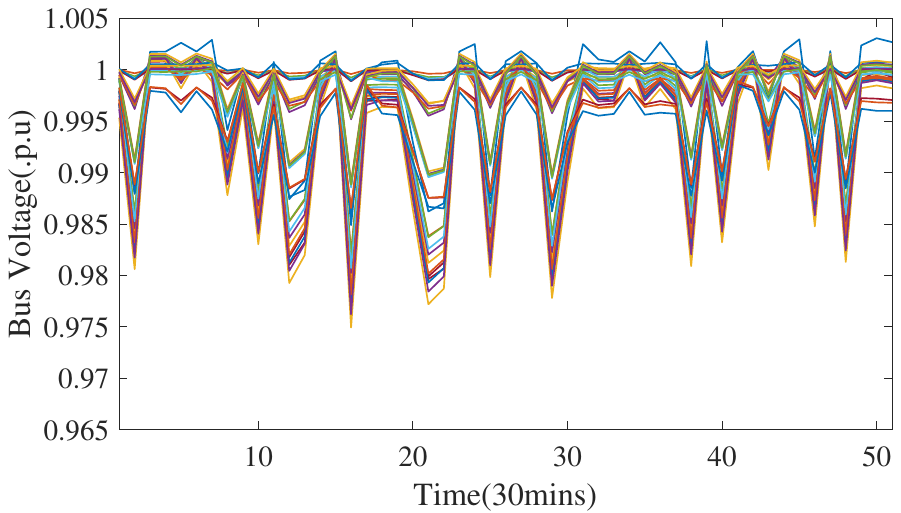}
		\end{minipage}
	}%
        \subfigure[C2]{
        \label{busvoltageDAC}
		\begin{minipage}[t]{0.46\linewidth}
			\centering
                 \vspace{-4.5cm}
			\includegraphics[width=1.8in]{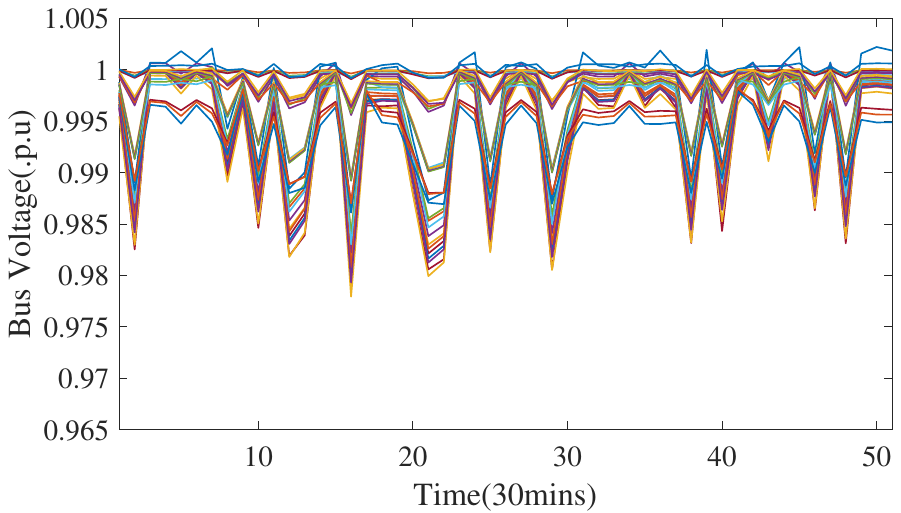}
		\end{minipage}
	}\\%
	\subfigure[C1]{
        \label{busvoltagereal}
		\begin{minipage}[t]{0.46\linewidth}
			\centering
                 \vspace{-4.5cm}
			\includegraphics[width=1.8in]{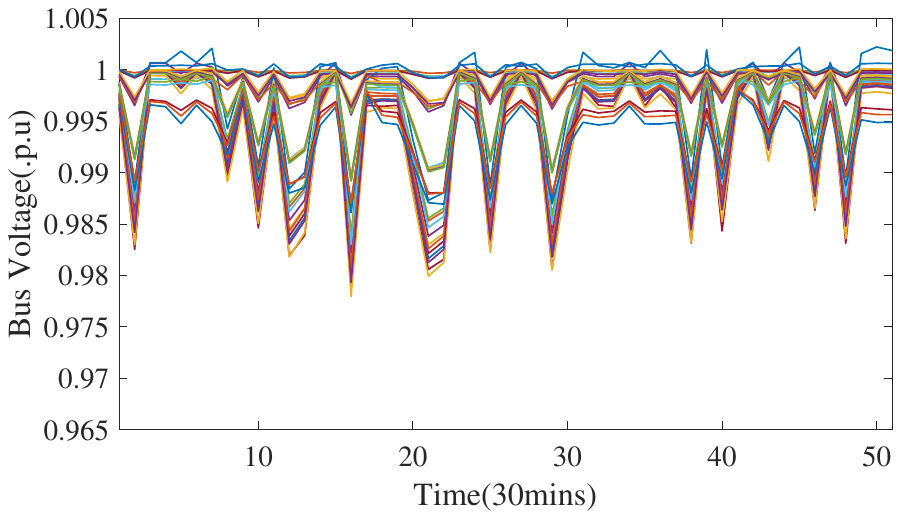}
		\end{minipage}
	}
        \subfigure[Proposed]{
        \label{nodevoltagepv11_RSMATD3_33}
		\begin{minipage}[t]{0.46\linewidth}
			\centering
                 \vspace{-4.5cm}
			\includegraphics[width=1.8in]{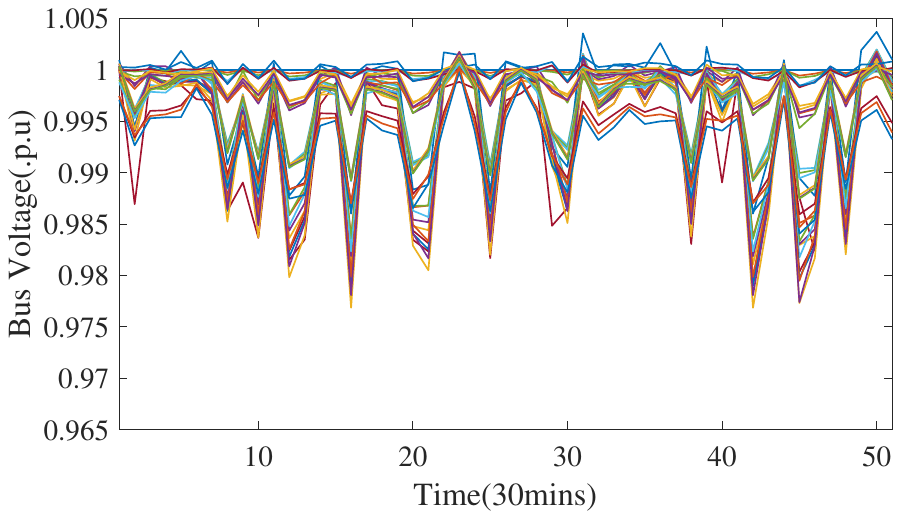}
		\end{minipage}
	}\\%
        \subfigure[C4]{
        \label{nodevoltagepv11_MATD3_33}
		\begin{minipage}[t]{0.46\linewidth}
			\centering
                 \vspace{-4.5cm}
			\includegraphics[width=1.8in]{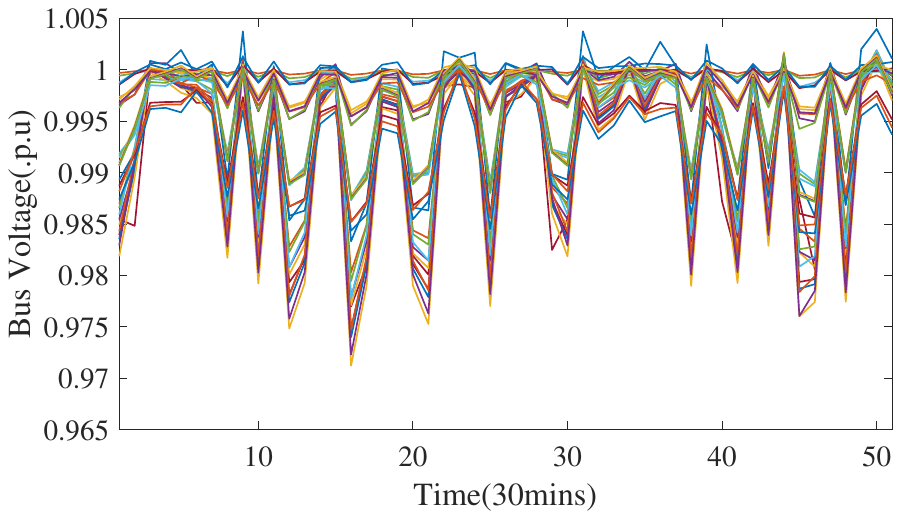}
		\end{minipage}
	}%
	\subfigure[C5]{
        \label{nodevoltagepv11_MADDPG_33}
		\begin{minipage}[t]{0.46\linewidth}
			\centering
                 \vspace{-4.5cm}
			\includegraphics[width=1.8in]{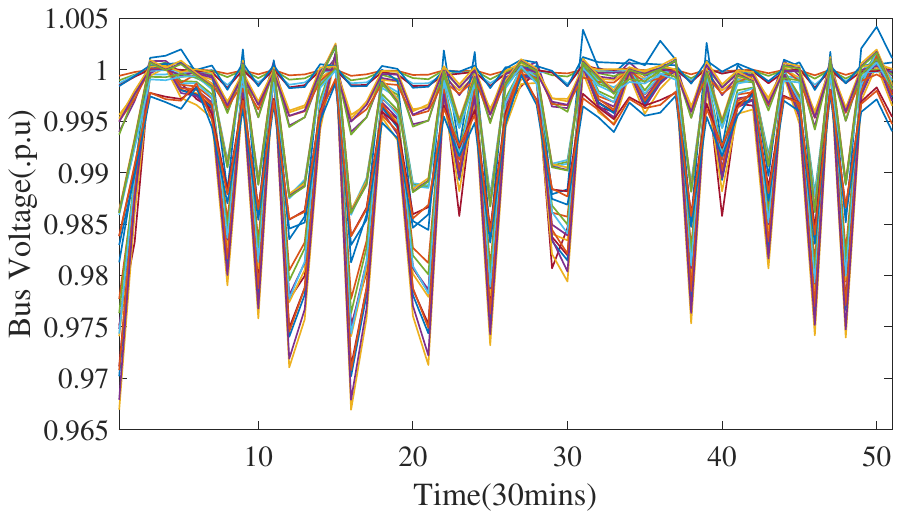}
		\end{minipage}
	}
	\centering
	\caption{Voltage variation of all 33 buses.}
	\label{fig_nodevoltage_bus33}
\end{figure}

Through VVC simulations using the test dataset, Fig. \ref{fig_nodevoltage_bus33} shows the bus voltage variation across different methods. Tab. \ref{devia_loss_bus33} lists the average total voltage deviation, average network loss, and maximum bus voltage deviation. In comparison to the scenario without voltage control, both C1-C5 and our proposed method can regulate the bus voltage within a safe range and effectively reduce network losses. 
From Fig. \ref{busvoltageNDAC10}-\ref{busvoltageDAC}, compared with the results in C3, the bus voltage in C2 consistently approaches the reference voltage, and in Tab. \ref{devia_loss_bus33}, the network loss and total voltage deviation of C2 inferior to those in C3. 
Combining Fig. \ref{busvoltagereal}, C2, leveraging our proposed DA method, demonstrates performance on par with C1.
The DA method utilizes the probability distribution of delay to mitigate inaccuracies in PV reactive power caused by prediction errors.

\begin{table}[!htb]  \centering
\caption{Average optimization target values and maximum bus voltage deviation of 33-bus System.}
\label{devia_loss_bus33}
\small
\begin{tabular}{|c|c|c|c|c|c|}
\hline
\multicolumn{2}{|c|}{ \multirow{2}*{\textbf{Methods}} }&\textbf{AverVol}&\textbf{AverPow}&\textbf{AverObj}&\textbf{MaxVol}\\
\multicolumn{2}{|c|}{}&\textbf{Devia/.p.u}&\textbf{Loss/MW}&\textbf{Value}&\textbf{Devia/.p.u}\\
\cline{1-6}    
\multicolumn{2}{|c|}{\textbf{No control}} & 1.2069 & 0.5091 & 0.858
 & 0.2893 \\ \hline  \multirow{3}*{\textbf{C3}}& \textbf{10s} & 0.141 & 0.27 & 0.205 & 0.027 \\ 
\cline{2-6}
& \textbf{5s} & 0.125 & 0.266 & 0.195 & 0.026 \\ 
\cline{2-6}
& \textbf{1s} & 0.113 & 0.261 & 0.187 & 0.025 \\ 
\cline{1-6}
\multicolumn{2}{|c|}{\textbf{C2}} & 0.112 & 0.256 & 0.1831 & 0.023\\ 
\cline{1-6}
\multicolumn{2}{|c|}{\textbf{C1}} & 0.11 & 0.256 & 0.183 & 0.022 \\ \hline
\multicolumn{2}{|c|}{\textbf{Proposed}} & 0.111 & 0.257 & 0.184 &  0.024  \\ \hline
\multicolumn{2}{|c|}{\textbf{C4}} & 0.147 & 0.276 & 0.212 & 0.029 \\ \hline
\multicolumn{2}{|c|}{\textbf{C5}} & 0.173 & 0.281 & 0.227 & 0.0289 \\ \hline

\end{tabular}
\end{table}

Fig. \ref{nodevoltagepv11_RSMATD3_33} and \ref{nodevoltagepv11_MADDPG_33} illustrate that the differences in bus voltages among our proposed method, C4, and C5 appear to be insignificant.
As detailed in Tab. \ref{devia_loss_bus33}, compared with C4 and C5, the max bus voltage deviation of our proposed method is the smallest. The average voltage deviation of our proposed method is 24.8\% and 36\% less than C4 and C5. 
Regarding average network loss, our proposed method is 6.8\% and 8.6\% smaller than theirs. 
The RS mechanism in our proposed algorithm ensures equal consideration of network loss and voltage deviation in the reward, preventing the neglect of one aspect and deviations from the optimal solution. 
Although our proposed method is inferior to C1 and C2, the optimal algorithm in them requires complete system topology model and parameters, resulting in high computational costs. And the calculation time of C1 and C2 is far greater than our proposed method.


\subsection{Robustness verification}

\begin{figure}
    \subfigure[Average Voltage]{
    \label{Avervoldevia_robust}
    \begin{minipage}[t]{0.46\linewidth}
        \centering
         \vspace{-4.5cm}
        \includegraphics[width=1.8in]{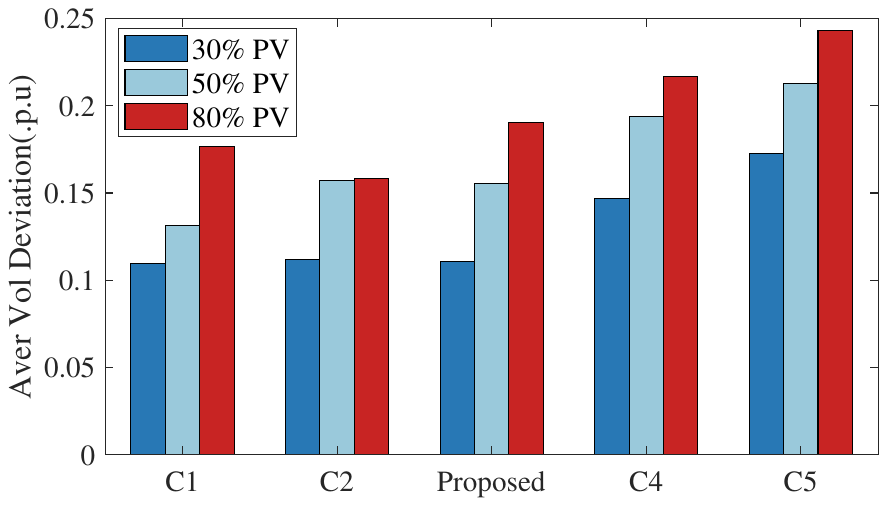}
    \end{minipage}%
}
    \subfigure[Maximum Voltage Deviation]{
    \label{Avermaxvoldevia_robust}
    \begin{minipage}[t]{0.46\linewidth}
        \centering
         \vspace{-4.5cm}
        \includegraphics[width=1.8in]{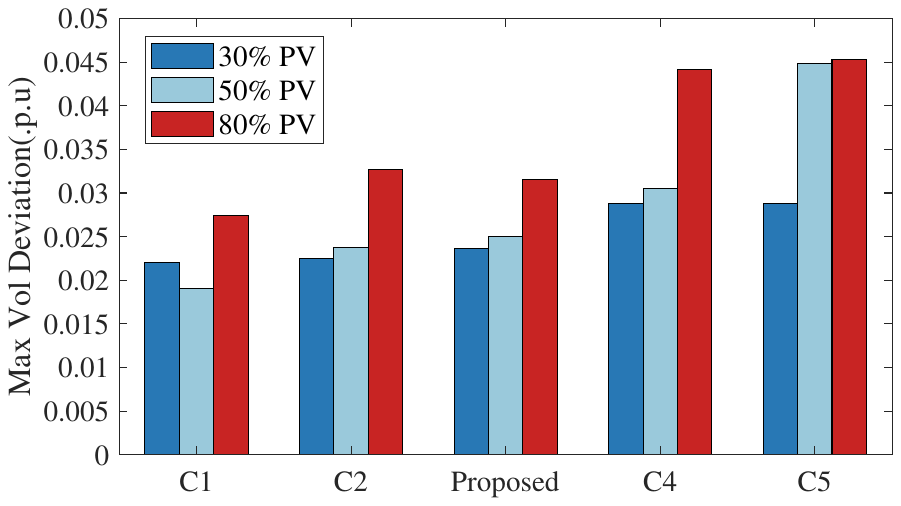}
     \end{minipage}
}
    \caption{The system performance over different PV penetration rates.}
    \label{fig_Verify_Robustness}
\end{figure}

We evaluate the robustness of our proposed method in managing the volatility of PV generation within ADNs. Fig. \ref{fig_Verify_Robustness} presents the results across various PV penetration levels
As depicted in Fig. \ref{Avervoldevia_robust}, an escalation in the PV penetration rate corresponds to an increase in the average total voltage deviation.
The incorporation of additional PV inverters leads to an excess of power generation, requiring the transmission of power to loads in other buses.
Consequently, the bus voltages are elevated, and the network losses also increase attributed to branch impedance.
Fig. \ref{Avermaxvoldevia_robust} illustrates that, across all three scenarios, our proposed method effectively regulates voltage within a narrow, safe range. This observation indicates the capability of our method to achieve robust voltage control in ADNs amid escalating PV power generation.

\subsection{Convergence Analysis}

\begin{figure*}
	\centering
        \subfigure[Proposed]{
        \label{MPNRS-MATD3}
		\begin{minipage}[t]{0.3\linewidth}
			\centering
                 \vspace{-4.5cm}
			\includegraphics[width=2in]{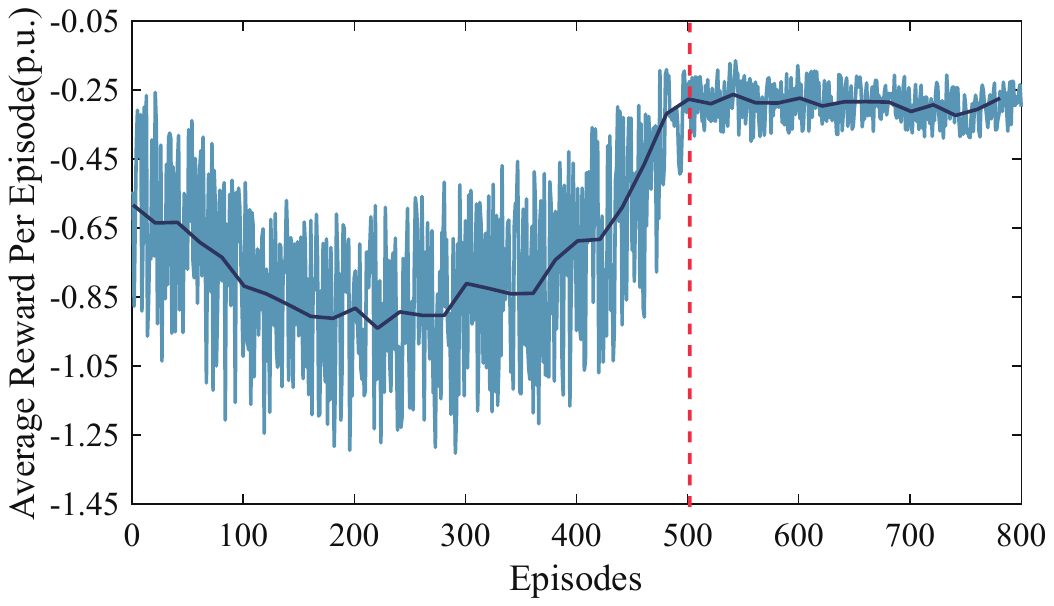}
		\end{minipage}
	}%
        \subfigure[C4]{
        \label{MATD3}
		\begin{minipage}[t]{0.3\linewidth}
			\centering
                 \vspace{-4.5cm}
			\includegraphics[width=2in]{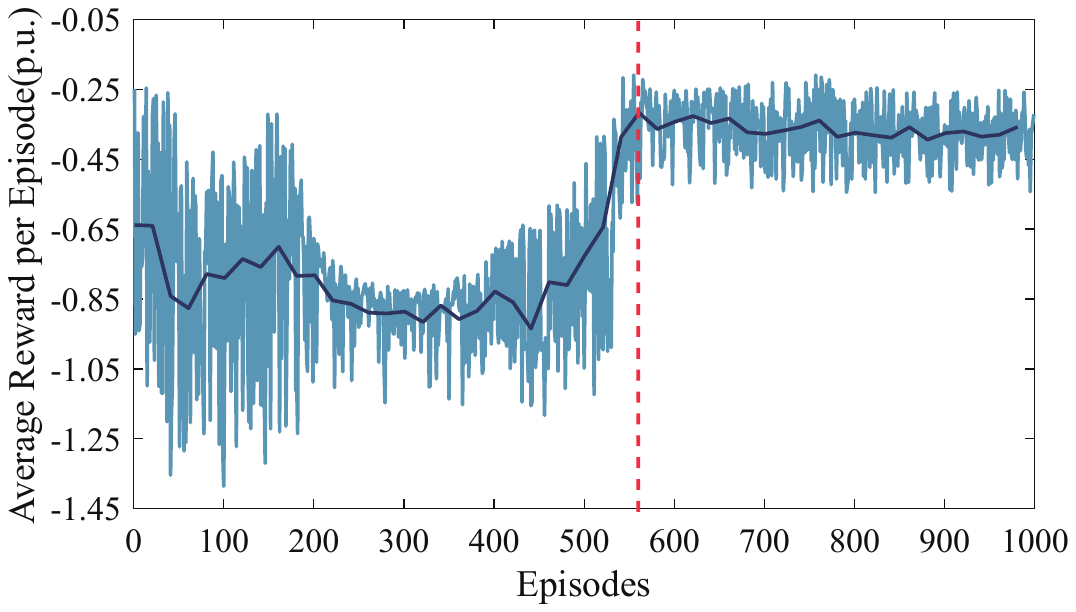}
		\end{minipage}
	}%
	\subfigure[C5]{
        \label{MADDPG}
		\begin{minipage}[t]{0.3\linewidth}
			\centering
                 \vspace{-4.5cm}
			\includegraphics[width=2in]{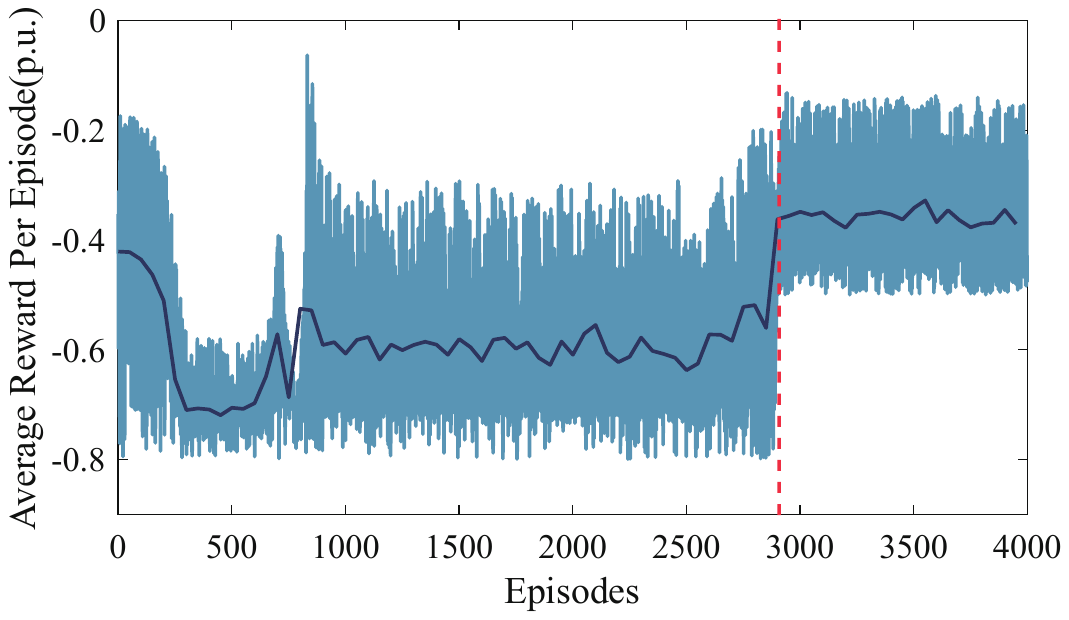}
		\end{minipage}
	}%
	\centering
	\caption{Comparison of convergence speed of three different algorithms.}
	\label{fig_Convergence_Speed}
\end{figure*}

To compare the convergence rate of the MARL algorithm (MPNRS-MATD3) in our proposed method, Fig. \ref{fig_Convergence_Speed} compares the changes in average rewards per episode of our proposed algorithm, C4 and C5 during the training process. 
Notably, the average reward exhibits convergence at approximately 3000 episodes for the C5 method, about 600 episodes for C4, and a notably swifter convergence at only 500 episodes for our proposed algorithm.
Moreover, the average reward fluctuation amplitude after the convergence of the MPNRS-MATD3 algorithm is slightly less than that of C4 and significantly less than in C5. 
In comparison to C4 and C5, our proposed MPNRS-MATD3 algorithm demonstrates superior convergence speed and enhanced training performance.
The RS mechanism within our proposed algorithm amplifies the impact of network loss and voltage deviation on the reward, consequently hastening the convergence speed.

\subsection{Scalability Proformance}

\begin{figure}[!htb]
   \centering
   \vspace{-9.5cm}
   \includegraphics[width=3.6in]{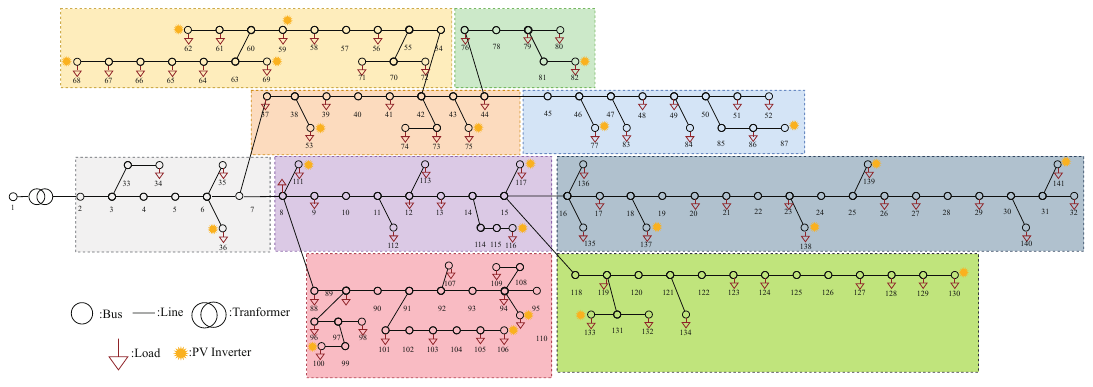}
   \caption{141-bus Network Topology.} \label{fig_IEEE141_Network_Topolopy}
\end{figure}

\begin{figure}
    \subfigure[Power Loss]{
    \label{power_loss_141}
    \begin{minipage}[t]{0.46\linewidth}
        \centering
         \vspace{-4.5cm}
        \includegraphics[width=1.8in]{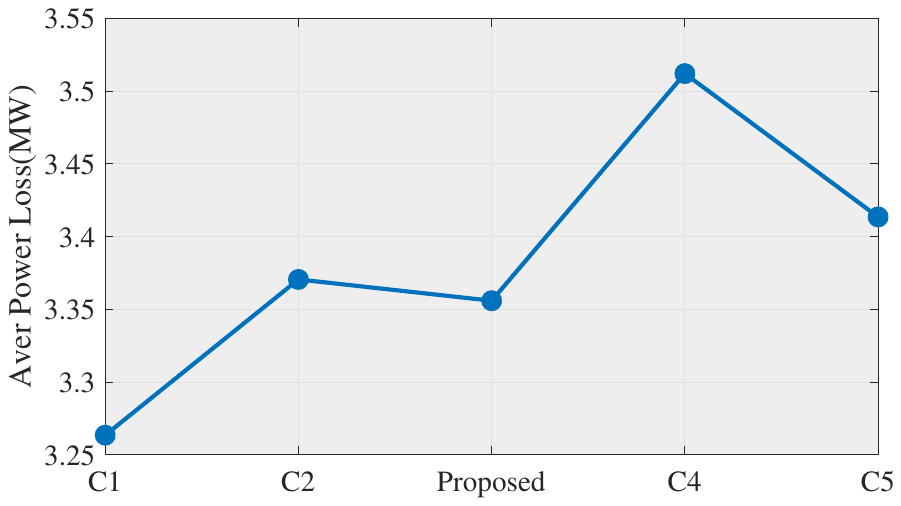}
    \end{minipage}%
}
    \subfigure[Voltage Deviation]{
    \label{vol_devia_141}
    \begin{minipage}[t]{0.46\linewidth}
        \centering
         \vspace{-4.5cm}
        \includegraphics[width=1.8in]{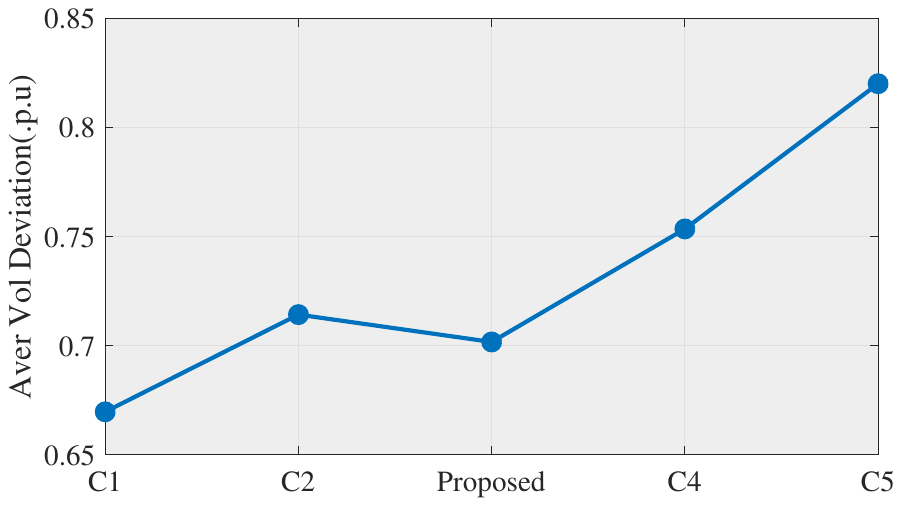}
     \end{minipage}
}\\
    \subfigure[Obj Value]{
    \label{obj_value_141}
    \begin{minipage}[t]{0.46\linewidth}
        \centering
         \vspace{-4.5cm}
        \includegraphics[width=1.8in]{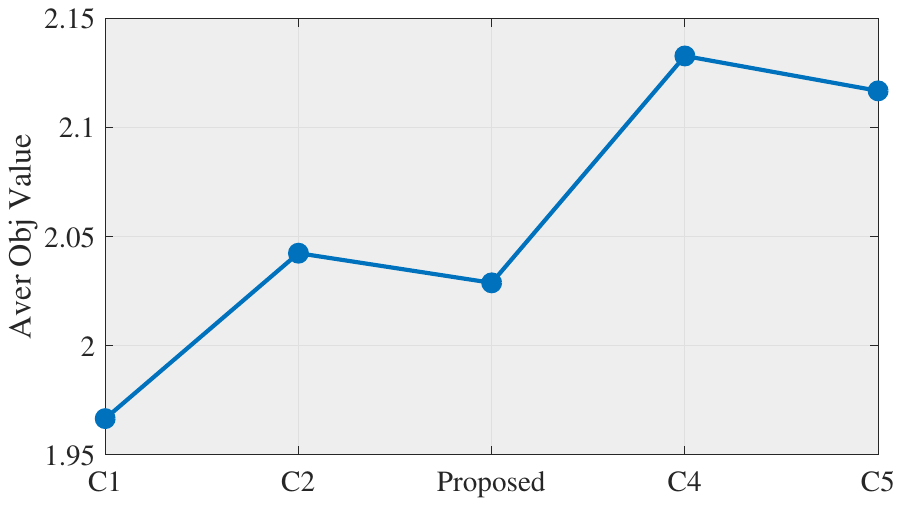}
     \end{minipage}%
}
    \subfigure[Voltage Distribution]{
    \label{voltage_distribution_141}
    \begin{minipage}[t]{0.46\linewidth}
        \centering
         \vspace{-4.5cm}
        \includegraphics[width=1.8in]{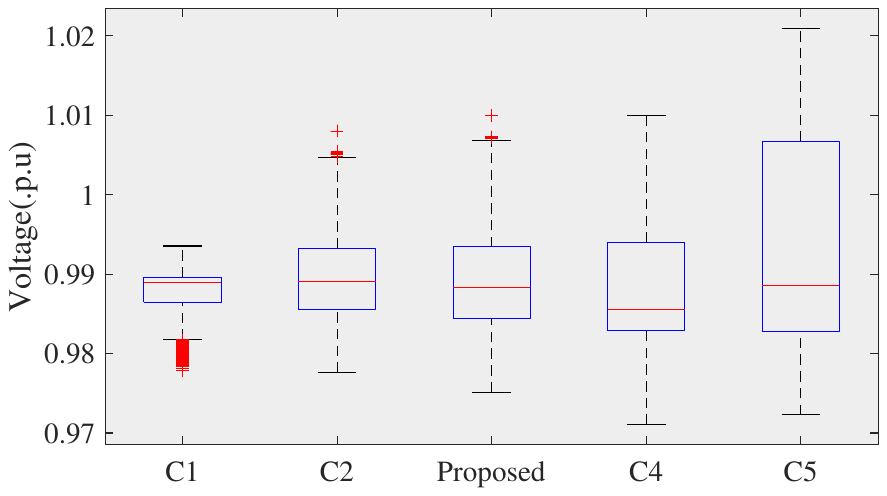}
    \end{minipage}
}
    \caption{The system performance of the 141-Bus network.}
    \label{fig_141_comparasion}
\end{figure}

To verify the scalability performance of our proposed scheme, simulations are conducted on a 141-bus network, as illustrated in Fig. \ref{fig_IEEE141_Network_Topolopy}, with partition regions detailed in \cite{wang2021multi}. 
The other parameters are shown in Tab. \ref{table_system_parameter}. Fig. \ref{fig_141_comparasion} presents the results of the average network power loss, average total voltage deviation, average objective value, and bus voltage distribution. 
In Fig. \ref{power_loss_141}-\ref{obj_value_141}, our proposed method exhibits performance inferior to C1 and C2 in the 141-bus network but outperforms C4 and C5. 
In Fig. \ref{voltage_distribution_141}, the bus voltage distribution range is smaller than C4 and C5. It indicates that our proposed MPNRS-MATD3 algorithm has near-optimal control performance and better scalability compared to the other two MARL-based competitors. 

\section{Conclusion}
\label{sec_conclusion}

This paper proposes a delay adaptive VVC framework for ADNs. The framework analyzes the probability distribution of imprecisely known system delay, mitigates the impact of state gap, and performs multiple system operation state prediction results to achieve the delay adaptive characteristics in voltage control. 
Additionally, by identifying the worst-performing system operation state through sample selection, the robust VVC problem ensures the robustness of voltage control. 
Finally, the Dec-POMDP model is used to reformulate the problem, and an MPNRS-MATD3 algorithm is designed to rapidly solve the problem. 
Simulation results show that the proposed framework successfully implements delay adaptive voltage control, and the control commands based on the proposed robust optimization problem and solving algorithm, demonstrate a high level of robustness in performance.




\ifCLASSOPTIONcaptionsoff
  \newpage
\fi


\section*{Acknowledgment}
This work is supported in part by grants from the National Natural Science Foundation of China (52077049, 62173120), the Anhui Provincial Natural Science Foundation (2008085UD04, 2108085UD07, 2108085UD11), the 111 Project (BP0719039).


\begin{thebibliography}{10}
\providecommand{\url}[1]{#1}
\csname url@samestyle\endcsname
\providecommand{\newblock}{\relax}
\providecommand{\bibinfo}[2]{#2}
\providecommand{\BIBentrySTDinterwordspacing}{\spaceskip=0pt\relax}
\providecommand{\BIBentryALTinterwordstretchfactor}{4}
\providecommand{\BIBentryALTinterwordspacing}{\spaceskip=\fontdimen2\font plus
\BIBentryALTinterwordstretchfactor\fontdimen3\font minus
  \fontdimen4\font\relax}
\providecommand{\BIBforeignlanguage}[2]{{%
\expandafter\ifx\csname l@#1\endcsname\relax
\typeout{** WARNING: IEEEtran.bst: No hyphenation pattern has been}%
\typeout{** loaded for the language `#1'. Using the pattern for}%
\typeout{** the default language instead.}%
\else
\language=\csname l@#1\endcsname
\fi
#2}}
\providecommand{\BIBdecl}{\relax}
\BIBdecl

\bibitem{abad2020PV}
M.~S.~S. Abad and J.~Ma, ``Photovoltaic hosting capacity sensitivity to active
  distribution network management,'' \emph{IEEE Transactions on Power Systems},
  vol.~36, no.~1, pp. 107--117, 2020.

\bibitem{zhao2017network}
B.~Zhao, Z.~Xu, C.~Xu, C.~Wang, and F.~Lin, ``Network partition-based zonal
  voltage control for distribution networks with distributed pv systems,''
  \emph{IEEE Transactions on Smart Grid}, vol.~9, no.~5, pp. 4087--4098, 2017.

\bibitem{generation2020ieee}
D.~Generation and E.~Storage, ``Ieee standard for interconnection and
  interoperability of distributed energy resources with associated electric
  power systems interfaces amendment 1: To provide more,'' \emph{IEEE:
  Piscataway, NJ, USA}, 2020.

\bibitem{PV2016}
F.~Ding, A.~Nagarajan, S.~Chakraborty, M.~Baggu, A.~Nguyen, S.~Walinga,
  M.~McCarty, and F.~Bell, ``Photovoltaic impact assessment of smart inverter
  volt-var control on distribution system conservation voltage reduction and
  power quality,'' National Renewable Energy Lab.(NREL), Golden, CO (United
  States), 2016.

\bibitem{2016POMDPs}
F.~A. Oliehoek and C.~Amato, ``A concise introduction to decentralized
  pomdps,'' \emph{Springer Publishing Company, Incorporated}, 2016.

\bibitem{RS1999policy}
A.~Y. Ng, D.~Harada, and S.~Russell, ``Policy invariance under reward
  transformations: Theory and application to reward shaping,'' in \emph{Icml},
  vol.~99.\hskip 1em plus 0.5em minus 0.4em\relax Citeseer, 1999, pp. 278--287.

\bibitem{lu2022reward}
R.~Lu, Z.~Jiang, H.~Wu, Y.~Ding, D.~Wang, and H.-T. Zhang, ``Reward
  shaping-based actor--critic deep reinforcement learning for residential
  energy management,'' \emph{IEEE Transactions on Industrial Informatics},
  vol.~19, no.~3, pp. 2662--2673, 2022.

\bibitem{turitsyn2011options}
K.~Turitsyn, P.~Sulc, S.~Backhaus, and M.~Chertkov, ``Options for control of
  reactive power by distributed photovoltaic generators,'' \emph{Proceedings of
  the IEEE}, vol.~99, no.~6, pp. 1063--1073, 2011.

\bibitem{David2020DeepAR}
J.~G. David~Salinas, Valentin~Flunkert and T.~Januschowski, ``Deepar:
  Probabilistic forecasting with autoregressive recurrent networks,''
  \emph{International Journal of Forecasting}, vol.~36, no.~3, pp. 1181--1191,
  2020.

\bibitem{Review2019Sun}
H.~Sun, Q.~Guo, J.~Qi, V.~Ajjarapu, R.~Bravo, J.~Chow, Z.~Li, R.~Moghe,
  E.~Nasr-Azadani, U.~Tamrakar, G.~N. Taranto, R.~Tonkoski, G.~Valverde, Q.~Wu,
  and G.~Yang, ``Review of challenges and research opportunities for voltage
  control in smart grids,'' \emph{IEEE Transactions on Power Systems}, vol.~34,
  no.~4, pp. 2790--2801, 2019.

\bibitem{Richardson2012}
P.~Richardson, D.~Flynn, and A.~Keane, ``Local versus centralized charging
  strategies for electric vehicles in low voltage distribution systems,''
  \emph{IEEE Transactions on Smart Grid}, vol.~3, no.~2, pp. 1020--1028, 2012.

\bibitem{Anton2017}
K.~E. Antoniadou-Plytaria, I.~N. Kouveliotis-Lysikatos, P.~S. Georgilakis, and
  N.~D. Hatziargyriou, ``Distributed and decentralized voltage control of smart
  distribution networks: Models, methods, and future research,'' \emph{IEEE
  Transactions on Smart Grid}, vol.~8, no.~6, pp. 2999--3008, 2017.

\bibitem{Zhang2023}
Z.~Zhang, C.~Dou, D.~Yue, Y.~Zhang, B.~Zhang, and B.~Li, ``Regional coordinated
  voltage regulation in active distribution networks with pv-bess,'' \emph{IEEE
  Transactions on Circuits and Systems II: Express Briefs}, vol.~70, no.~2, pp.
  596--600, 2023.

\bibitem{Safe2020}
W.~Wang, N.~Yu, Y.~Gao, and J.~Shi, ``Safe off-policy deep reinforcement
  learning algorithm for volt-var control in power distribution systems,''
  \emph{IEEE Transactions on Smart Grid}, vol.~11, no.~4, pp. 3008--3018, 2020.

\bibitem{Jafari2018}
M.~Jafari, T.~O. Olowu, and A.~I. Sarwat, ``Optimal smart inverters volt-var
  curve selection with a multi-objective volt-var optimization using
  evolutionary algorithm approach,'' in \emph{2018 North American Power
  Symposium (NAPS)}, 2018, pp. 1--6.

\bibitem{Liu2021Robust}
H.~Liu, C.~Zhang, Q.~Chai, K.~Meng, Q.~Guo, and Z.~Y. Dong, ``Robust regional
  coordination of inverter-based volt/var control via multi-agent deep
  reinforcement learning,'' \emph{IEEE Transactions on Smart Grid}, vol.~12,
  no.~6, pp. 5420--5433, 2021.

\bibitem{Zhang2022MATD3}
B.~Zhang, Z.~Chen, X.~Wu, D.~Cao, and W.~Hu, ``A matd3 -based voltage control
  strategy for distribution networks considering active and reactive power
  adjustment costs,'' in \emph{2022 IEEE International Conference on Power
  Systems and Electrical Technology (PSET)}, 2022, pp. 189--194.

\bibitem{brown2017pypsa}
T.~Brown, J.~H{\"o}rsch, and D.~Schlachtberger, ``Pypsa: Python for power
  system analysis,'' \emph{arXiv preprint arXiv:1707.09913}, 2017.

\bibitem{zimmerman1997matpower}
R.~D. Zimmerman, C.~E. Murillo-S{\'a}nchez, and D.~Gan, ``Matpower,''
  \emph{PSERC.[Online]. Software Available at: http://www. pserc. cornell.
  edu/matpower}, 1997.

\bibitem{Wang2020Data}
S.~Wang, J.~Duan, D.~Shi, C.~Xu, H.~Li, R.~Diao, and Z.~Wang, ``A data-driven
  multi-agent autonomous voltage control framework using deep reinforcement
  learning,'' \emph{IEEE Transactions on Power Systems}, vol.~35, no.~6, pp.
  4644--4654, 2020.

\bibitem{Two2021Sun}
X.~Sun and J.~Qiu, ``Two-stage volt/var control in active distribution networks
  with multi-agent deep reinforcement learning method,'' \emph{IEEE
  Transactions on Smart Grid}, vol.~12, no.~4, pp. 2903--2912, 2021.

\bibitem{Attention2021Cao}
D.~Cao, J.~Zhao, W.~Hu, F.~Ding, Q.~Huang, and Z.~Chen, ``Attention enabled
  multi-agent drl for decentralized volt-var control of active distribution
  system using pv inverters and svcs,'' \emph{IEEE Transactions on Sustainable
  Energy}, vol.~12, no.~3, pp. 1582--1592, 2021.

\bibitem{Deep2022Cao}
D.~Cao, J.~Zhao, W.~Hu, N.~Yu, F.~Ding, Q.~Huang, and Z.~Chen, ``Deep
  reinforcement learning enabled physical-model-free two-timescale voltage
  control method for active distribution systems,'' \emph{IEEE Transactions on
  Smart Grid}, vol.~13, no.~1, pp. 149--165, 2022.

\bibitem{Mean2021Wei}
B.~Wei, Z.~Qiu, and G.~Deconinck, ``A mean-field voltage control approach for
  active distribution networks with uncertainties,'' \emph{IEEE Transactions on
  Smart Grid}, vol.~12, no.~2, pp. 1455--1466, 2021.

\bibitem{Gorbachev2023Distributed}
S.~Gorbachev, A.~Mani, L.~Li, L.~Li, and Y.~Zhang, ``Distributed energy
  resources based two-layer delay-independent voltage coordinated control in
  active distribution network,'' \emph{IEEE Transactions on Industrial
  Informatics}, pp. 1--10, 2023.

\bibitem{Xing2022}
L.~Xing, Y.~Mishra, Y.-C. Tian, G.~Ledwich, C.~Wen, W.~He, W.~Du, and F.~Qian,
  ``Distributed voltage regulation for low-voltage and high-pv-penetration
  networks with battery energy storage systems subject to communication
  delay,'' \emph{IEEE Transactions on Control Systems Technology}, vol.~30,
  no.~1, pp. 426--433, 2022.

\bibitem{Sang2023Safety}
L.~Sang, Y.~Xu, H.~Long, and W.~Wu, ``Safety-aware semi-end-to-end coordinated
  decision model for voltage regulation in active distribution network,''
  \emph{IEEE Transactions on Smart Grid}, vol.~14, no.~3, pp. 1814--1826, 2023.

\bibitem{Sizing2021}
A.~F. Nematollahi, H.~Shahinzadeh, H.~Nafisi, B.~Vahidi, Y.~Amirat, and
  M.~Benbouzid, ``Sizing and sitting of ders in active distribution networks
  incorporating load prevailing uncertainties using probabilistic approaches,''
  \emph{Applied Sciences}, vol.~11, no.~9, 2021.

\bibitem{wang2021multi}
J.~Wang, W.~Xu, Y.~Gu, W.~Song, and T.~C. Green, ``Multi-agent reinforcement
  learning for active voltage control on power distribution networks,''
  \emph{Advances in Neural Information Processing Systems}, vol.~34, pp.
  3271--3284, 2021.

\bibitem{Jain2019Quasi}
A.~K. Jain, K.~Horowitz, F.~Ding, N.~Gensollen, B.~Mather, and B.~Palmintier,
  ``Quasi-static time-series pv hosting capacity methodology and metrics,'' in
  \emph{2019 IEEE Power \& Energy Society Innovative Smart Grid Technologies
  Conference (ISGT)}, 2019, pp. 1--5.

\bibitem{Gor2023Distr}
S.~Gorbachev, A.~Mani, L.~Li, L.~Li, and Y.~Zhang, ``Distributed energy
  resources based two-layer delay-independent voltage coordinated control in
  active distribution network,'' \emph{IEEE Transactions on Industrial
  Informatics}, pp. 1--10, 2023.

\bibitem{Gholami2020Robust}
M.~Gholami, A.~Pisano, and E.~Usai, ``Robust distributed optimal secondary
  voltage control in islanded microgrids with time-varying multiple delays,''
  in \emph{2020 IEEE 21st Workshop on Control and Modeling for Power
  Electronics (COMPEL)}, 2020, pp. 1--8.

\end{thebibliography}




\end{document}